\documentclass[11pt]{article}
\usepackage{authblk}
\usepackage{graphicx}
\usepackage{color, colortbl}
\usepackage[caption=false,font=footnotesize]{subfig}
\usepackage{amsmath, amssymb}
\usepackage{enumitem}
\usepackage{multirow}

\usepackage{float}

\usepackage{lineno}

\begin{document}

\title{A Statistical Density-Based Analysis of Graph Clustering Algorithm Performance}

\author[1]{Pierre Miasnikof \thanks{corresponding author: p.miasnikof@mail.utoronto.ca}}
\author[2]{Alexander Y. Shestopaloff}
\author[1]{Anthony J. Bonner}
\author[1]{Yuri Lawryshyn}
\author[3,4]{Panos M. Pardalos}

\affil[1]{University of Toronto, Toronto, ON, Canada}
\affil[2]{The Alan Turing Institute, London, United Kingdom}
\affil[3]{University of Florida, Gainesville, FL, USA}
\affil[4]{National Research University HSE, Nizhny Novgorod, Russian Federation}

\date{}

\maketitle

{ \bf \color{blue} NOTE: A slightly modified version of this article, titled ``A Density-Based Statistical Analysis of Graph Clustering Algorithm Performance'', has been accepted for publication in the Journal of Complex Networks, published by Oxford University Press. }

\begin{abstract}
Measuring graph clustering quality remains an open problem. To address it, we introduce quality measures based on comparisons of global, intra- and inter-cluster densities, an accompanying statistical significance test and a step-by-step routine for clustering quality assessment. In doing so, we also offer our own definition of good clustering, as well as necessary and sufficient conditions that characterize it. Our null model does not rely on any generative model for the graph, unlike modularity which uses the configuration model as a null. Our measures are also shown to meet the axioms of a good clustering quality function, unlike the very commonly used modularity measure. They also have an intuitive graph-theoretic interpretation, a formal statistical interpretation and can be easily tested for significance. Our work is centered on the idea that well clustered graphs will display a mean intra-cluster density that is higher than global density and mean inter-cluster density. We develop tests to verify the existence of such a cluster structure. We empirically explore the behavior of our measures under a number of stress test scenarios and compare their behavior to the commonly used modularity and conductance measures. Empirical stress test results confirm that our measures compare very favorably to the established ones. In particular, they are shown to be more responsive to graph structure, less likely to breakdown during numerical implementation, less sensitive to uncertainty in connectivity and consistent with the axioms defining good clustering quality functions. These features are especially important in the context of larger data sets or when the data may contain errors in the connectivity patterns.
\end{abstract}

\section{Introduction}
While there are many graph clustering \footnote{Note on vocabulary: Although there are subtle differences between the concepts of graph clustering and community detection, we use the two interchangeably.} algorithms in the literature (e.g., \cite{NewGirvOrig2004,FanPard2010CND,FanPard2010LQ,highQuality2012,AloiseEtAl2012,GRASP_Pitsoulis}), measuring their performance, assessing the quality of the clusters they identify, remains an open problem \cite{BenchmarkLanchinetti,LanFor2009,empiricalLeskovec2010,bestmetric,Moschopoulos2011,YangLesko2012,Laar14,Handbook2015,qualityFns2016,MetricsScale16,Biswas2017,sizeMatters17,Pitsoulis2018,LiudaSynthetic2019}. In fact, it's been stated in a very recent publication that {\it``in the literature, there is no universally accepted metric for evaluating the performance of community detection algorithms'' }\cite{LiudaSynthetic2019}.

Graph clustering is the process of assigning common labels to vertices that are considered similar, vertices that should belong to a common set (cluster). It is a form of unsupervised learning, where one typically cannot count on labeled data to assess results. For example, Reichardt and Bornholdt correctly assert that ``{\it (...) running a clustering algorithm over a set of randomly generated data points will always produce clusters which, however, have little meaning}'' \cite{trulyModReichardt2006} . For this reason, our only quality measure is a thorough examination of the graph's and resulting clusters' connectivity patterns.

In this article, we present new (algorithm-independent) clustering performance measures to assess the strength of the clustering returned by any algorithm. Our measures can also provide comparisons of several clustering algorithms on a given graph. Through this presentation, we also offer our own definition of clustering quality and techniques to test its statistical significance. 

Our measures are based on comparisons of global, intra- and inter-cluster densities. Our null model does not rely on any generative model for the graph, unlike modularity which uses the configuration model as a null. Also, unlike the context-dependent approach presented by Creusefond et al. \cite{qualityFns2016}, we propose a general purpose quality measure based on graph and subgraph density, well known graph characteristics, and formal statistical testing. 

We begin with a review of two of the most common clustering performance measures, modularity and conductance. We empirically demonstrate how these measures may be drowned out by graph structure and lack sensitivity to it.  We also demonstrate how our test of clustering quality based on our two statistical measures of graph structure is more robust, easier to interpret and consistent with the axioms of good clustering. 

\begin{figure}
\centering
\subfloat[Well Clustered Graph]{ \includegraphics[width = 0.45\textwidth]{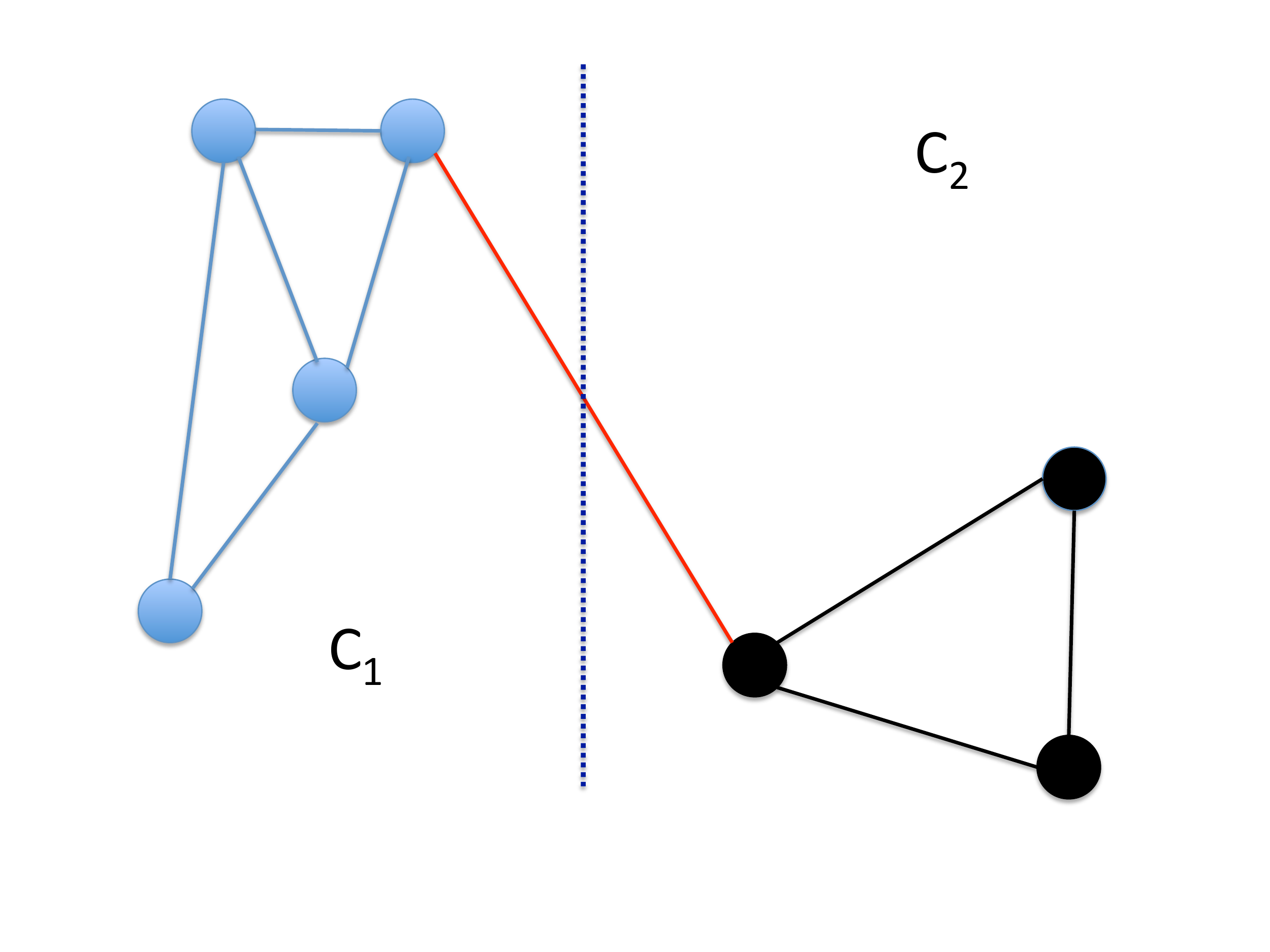}   \label{good} }
\subfloat[Improperly Clustered Graph]{ \includegraphics[width = 0.45\textwidth]{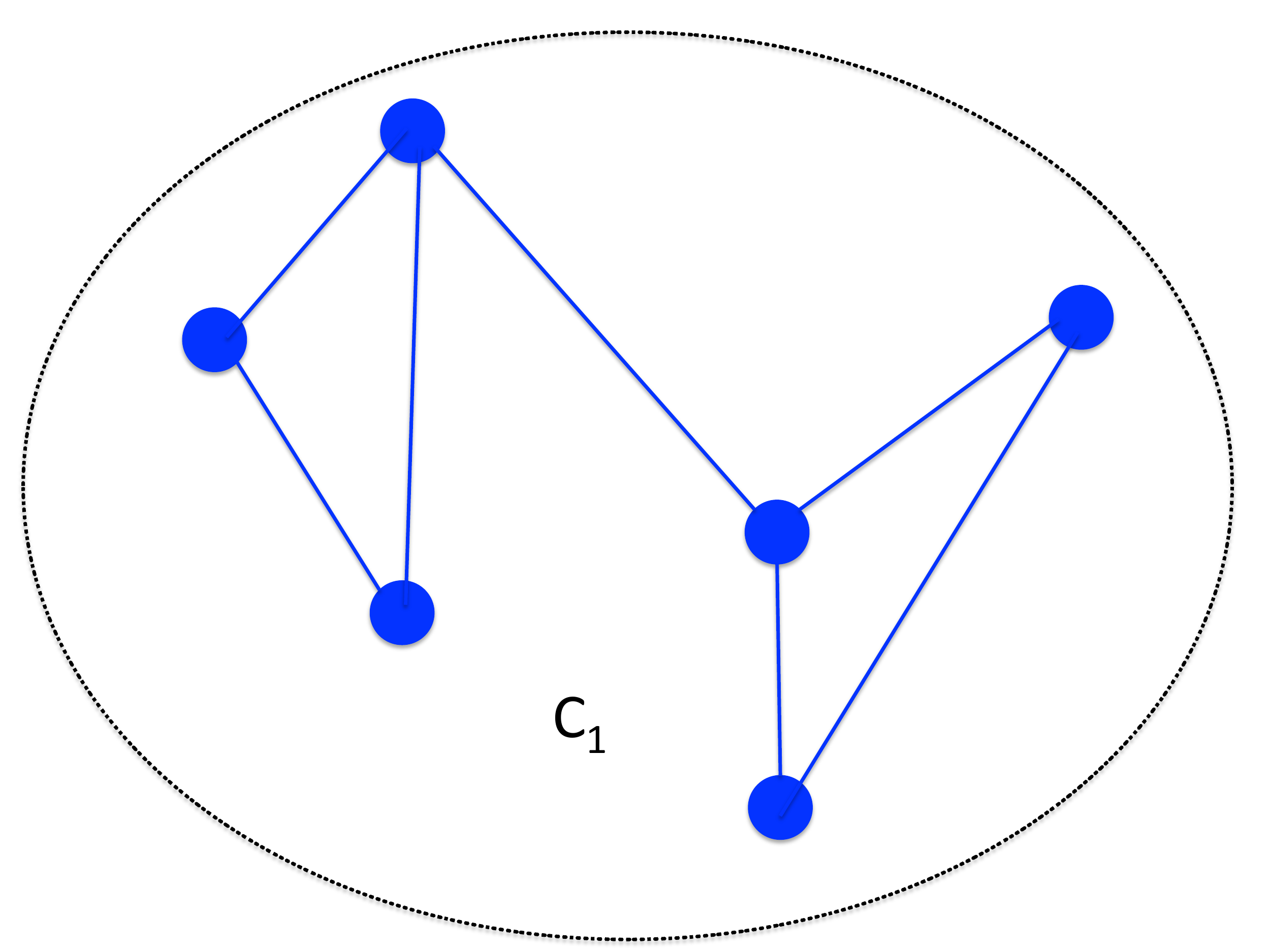} \label{nogood}  } \\
\caption{Examples of Good and Bad Clustering}
\label{goodNnogood}
\end{figure}

We restrict our attention to undirected unweighted and weighted graphs with positive edge weights. The graphs we consider also have no self-loops or multiple edges. It is important to emphasize that we are not trying to identify clusters and their constituent vertices, in this article. The work in this article focuses exclusively on assessing the quality of the clusters identified by a clustering algorithm. For example, we want an objective measure that allows us to conclude the algorithm that clustered the graph in Figure~\ref{good} performed well, while the algorithm that clustered the graph in Figure~\ref{nogood} performed poorly and did not partition the graph adequately. 

\section{Commonly Used Performance Measures, Quality Functions}
The  term ``clustering quality function'' of Van Laarhoven and Marchiori \cite{Laar14} is often used to designate clustering performance measures. These authors use the term ``quality function'' to describe a function that takes in a graph $G$ and a set of node clusters $C = \{c_1,c_2,\ldots\}$ and returns a real number, the quality measure. All the measures discussed in this article fit this description.

In this section, we describe the two most frequently used clustering quality functions, modularity and conductance. These functions are not only used to assess clustering quality but are also often used as objective functions in optimization-based clustering techniques (e.g., \cite{Louvain2008,Spielman2013}).

\subsection{Modularity}
Modularity ($Q$) is by far the most popular measure of clustering performance \cite{NewGirvOrig2004,ModClausetNewman2004,modBrandes2007,FortunatoLong2010,empiricalLeskovec2010,Kehagias2013,modWAW2016,EuroComb2017}. Originally introduced by Newman and Girvan in 2004 \cite{NewGirvOrig2004}, it has since been extensively used both as a performance measure and objective function to be maximized (e.g., \cite{NewGirvOrig2004,usingDjidjev2012,AloiseEtAl2012,GRASP_Pitsoulis}).  In this section, we present modularity $(Q)$ as shown in Clauset et al. \cite{ModClausetNewman2004}. Modularity is defined as 
\begin{eqnarray*}
Q &=& \sum_{i=1}^{k} \left( \underbrace{ e_{ii} - a_i^2 }_{ q_i } \right)  \\
\mbox{where,}  \\
e_{ii} &=& \frac{1}{2m} \sum_{v,w} A_{v,w} \, \delta (c_v, i) \delta (c_w, i) \\
a_i &=& \frac{1}{2m} \sum_v A_{v,.} \vec{1} \, \delta (c_v, i) \, .
\end{eqnarray*}
Here, $m= \vert E \vert $ is the total number of edges in the graph, $k$ is the number of clusters, $A_{v,w}$ is the element at the intersection of the $v$-th row and $w$-th column of the adjacency matrix, $A_{v,.}$ is the entire $v$-th row of the adjacency matrix, $\vec{1}$ is a vector of ones of appropriate dimension, $\delta(x,y)$ is the Kroenecker delta function and $c_v$ is the cluster in which vertex $v$ is clustered by the algorithm. Expanding our expression above using the graph's adjacency matrix, we get 
\begin{equation} 
Q = \sum_{i=1}^{k}  \left[ \underbrace{\frac{1}{2m} \sum_{v,w} A_{v,w} \, \delta (c_v, i) \delta (c_w, i)}_{e_{ii}} - \underbrace{\frac{1}{4m^2} \left( \sum_v A_{v,.} \vec{1} \, \delta (c_v, i) \right)^2}_{a_i^2}  \right] \, . \label{mod}
\end{equation}
Modularity always lies on the interval $[-\frac{1}{2},1]$ \cite{modBrandes2007}. Values greater than $0.3$ typically indicate a significant clustering \cite{ModClausetNewman2004}.  

In closing, it should be noted that modularity's biggest weakness is that it suffers from resolution limit. This weakness is well documented in the literature. For example, Fortunato and Bath\'{e}lemy \cite{ResolLimitFortunato2007} devote an entire article to this topic. These authors describe how any clustering quality function that is defined as a sum of qualities of individual clusters, where terms from smaller clusters are dominated by terms from larger clusters, suffers from resolution limit. Because the smaller clusters' contribution to the sum is dominated by the larger clusters', the final result is also dominated and does not always reflect structure accurately. Indeed, in Equation (\ref{mod}) we see how larger clusters dominate the outer summation. Modularity also suffers from many other degeneracies, as described by Good et al. \cite{GoodEtAl2010} among others. Additionally, Fortunato showed the difficulty of conducting statistical tests on modularity, due to the challenge of  identifying its true distribution \cite{FortunatoLong2010}. Finally, we note that Van Laarhoven and Marchiori have demonstrated modularity fails to meet some of the axioms of a good clustering quality function, namely locality and monotonicity \cite{Laar14}.

\subsection{Conductance}
Conductance ($\phi, \Phi$) is another popular clustering performance measure \cite{statpropLesko08,empiricalLeskovec2010,YangLesko2012,Spielman2013,qualityFns2016}. It is also used by some authors as an objective to be minimized (e.g.,  \cite{GoodBadSpectral2004,Spielman2013}). In this article, we use the definition presented by Spielman and Teng \cite{Spielman2013}. At the individual cluster level, conductance is defined as
\[
\phi(c_i) = \frac{ \partial (c_i) }{ \min \left( d(c_i), d(V \setminus c_i) \right) } \, .\label{phi}
\]
While at the graph level, it is defined as
\[
\Phi(G) = \min_{c_i} \phi(c_i) \, .
\]
Here, $\partial(c_i)$ is the number of edges joining vertices in cluster $c_i$ to vertices outside $c_i$, $d(c_i)$ is the sum of vertex degrees within $c_i$ and $d(V \setminus c_i)$ the sum of vertex degrees on the graph, outside $c_i$. A low conductance indicates strongly connected clusters. 

\section{A Definition of Good Clustering and New Clustering Performance Measures} \label{qualitydefn}
We now present our statistical performance measures, the Kappas. Before we begin, we introduce our own definition of good clustering. The Kappas are rooted in this definition. Our performance measures are based on a comparison of global, inter- and intra-cluster densities. We argue that the difference between these quantities is a measure of clustering strength.

\subsection{A Definition of Clustering Quality} 
In accordance with every other definition of a good clustering, we expect that an efficient clustering algorithm will label vertices such that intra-cluster connectivity is greater than global and inter-cluster connectivity (e.g., \cite{FortunatoLong2010,modWAW2016,EuroComb2017}) (if the graph does indeed have a clustered structure). We expect that nodes with the same label will have more common connections to each other than to ones with different labels. Alternately, we expect mostly uniform connectivity between vertices in the case of improperly clustered graphs. In summary, we expect that clustered vertices will form dense subgraphs within a sparser graph, when they are properly labeled. 

In step with these expectations, to gain a macroscopic view of the entire graph, we set up our average case benchmarks. We posit that a good clustering will group vertices so they form clusters whose vertices are more strongly connected to each other than to vertices belonging to different clusters, on average. To  gauge the strength of clustering, we rely on graph theory and statistics. We compare mean intra-cluster density, the graph's overall density and mean inter-cluster density. These quantities are defined below, in Section~\ref{kappas}. 

We postulate that a good clustering must meet two necessary conditions. Together, these necessary conditions form a sufficient condition for a clustering to be of good quality, on average. First, a good clustering should be characterized by a mean intra-cluster density that is higher than the graph's global density. Global density should, in turn, also be higher than mean inter-cluster density. Meeting this set of inequalities is the first necessary condition. 

The second necessary condition, also an inequality, is that mean intra-cluster density must be significantly greater than mean-inter cluster density. This last inequality automatically holds numerically, whenever the first necessary condition is met. However, this second necessary condition is a condition on statistical significance. We want the difference between these two mean densities, a quantity we call $\gamma$, to be non-trivial, statistically significant.  

The idea of using mean ``intra-connectivity'' and mean ``inter-connectivity'' to measure clustering strength was initially presented by Mancoridis et al. \cite{Mancoridis98}, although their definitions were not based on simple strict graph density. These authors' goal was to find optimal partitions of software system components and their model was based on directed graphs with self-loops. Later, Kannan et al. \cite{GoodBadSpectral2004} used global inter-cluster density as an indicator of poor clustering, to be minimized along with conductance. Fortunato \cite{FortunatoLong2010} then introduced the idea of using inter- and intra-cluster densities and comparing them to global density as a measure of the clustering strength of single clusters. We extend these ideas to gain a macroscopic view of the entire graph's cluster labeling, using the standard definition of density. 

While they are inspired by Mancoridis et al. and by Fortunato's follow-up on those ideas, our measures are tailored to undirected graphs and provide a macroscopic view. Unlike the measures proposed by Mancoridis et al., they correspond to standard definitions for simple graphs and can be generalized to weighted graphs as well. Unlike the measures proposed by Fortunato, they offer a graph level, not cluster level, picture. 

In the case of unweighted graphs, our measures are bounded within the interval $[0,1]$. In the case of weighted graphs, they become proportional to edge weight. High values denote densely connected graphs, clusters or cluster pairs and vice-versa. Also, because we take means over the entire graph, our measures provide a graph-wide picture, have statistical meaning and can easily be subjected to hypothesis testing. For these reasons, our measures are also more meaningful and more solidly grounded in graph and statistical theories than either conductance or modularity.

The well established and widely used measures of clustering strength, modularity and conductance, measure intra-cluster connectivity strength. Instead, we measure the strength of intra- and inter-cluster connectivity relative to each other and to the overall graph's connectivity. In doing so, we tailor our definition of and conclusions on the quality of clustering to the specific graph structure being analyzed. For example, in a moderately densely connected graph we expect clusters to be even more strongly connected. Meanwhile, moderately strong inter-cluster connections can be consistent with a good partition, as long as it remains weaker than intra-cluster connectivity. Conversely, in a sparsely connected graph, moderately strong inter-cluster connectivity is a symptom of a poor clustering. 

Our claim regarding inter- and intra-cluster densities is empirically validated by the simulations of our null hypothesis, shown in Section~\ref{nulls}. Through those simulations, we empirically confirm our hypothesis that in a random clustering of a graph, arguably the poorest of non-trivial clusterings, the difference between mean intra-cluster and mean inter-cluster densities is approximately zero. 

\subsection{The Kappas, Measures of Density} \label{kappas}
We define Kappa $(K)$ as the graph's overall connectivity measure, mean Kappa intra-cluster $(\bar{K}_{\mbox{intra}})$ as the measure of intra-cluster connectivity and  mean Kappa inter-cluster $(\bar{K}_{\mbox{inter}})$ as the measure of inter-cluster connectivity. These quantities are the graph's global density, mean intra-cluster density and mean inter-cluster density, respectively. Here, we extend Fortunato's  idea of examining inter- and intra-cluster density to determine the strength of a clustering \cite{FortunatoLong2010}, but with a focus on average-case subgraphs, not individual clusters. As mentioned in the previous section, we expect that under a good clustering the inequalities $\bar{K}_{\mbox{inter}} < K < \bar{K}_{\mbox{intra}}$ will hold. These inequalities are similar to those formulated by Fortunato \cite{FortunatoLong2010}, but they are formulated at a graph-wide scale, on the basis of graph-wide average-cases. Our model also allows these inequalities to be formulated as statistical hypothesis tests, as will be shown later. 

Below, we present the formulation for our clustering measures, for an unweighted undirected graph, but they easily generalize to weighted graphs as well. For weighted graphs, when edge weights are proportional to connection strength, our measures are computed by replacing the cardinality of edge sets (edge counts) with the sums of the corresponding edge weights (all edge weights or intra-/inter-cluster weights). However, it should be noted that densities in the weighted case are no longer contained within the interval $[0,1]$, although they remain non-negative (since we only consider positive edge weights).

In our formulation, we use the following variables. The set of all clusters is $C = \{ c_1,\ldots, c_\ell \}$, with $ \vert C \vert = \ell$, the total number of vertices in the graph is $\vert V \vert = N$, the total number of vertices in cluster $i$ is $n_i$, the set of all edges on the graph is $E = \{ e_1, \ldots, e_m \}$, where $\vert E \vert = m$. Finally,  $E_{ij}$ is the set of edges connecting a vertex in cluster $i$ to a vertex in cluster $j$, and $\vert E_{ij} \vert = m_{ij}$.  As a special case, note that $E_{ii}$ is the set of edges within cluster $i$, and $m_{ii}$ is the number of edges connecting vertices within cluster $i$.

Using standard definitions, we take the ratio of the edge counts (or sum of weights) over the maximum possible number of edges given the number of vertices. For mean intra- and inter-cluster connectivity, we compute the ratio for each cluster or pair of clusters and take their mean as a graph-wide measure.

We compute the graph's connections ratio, global density, as
\[
K = \frac{ \vert E \vert }{ 0.5 \times N (N-1) } \; .
 \]
This quantity is the ratio of the total number of edges over the number of edges in a complete graph with the same number of vertices. In the case of an unweighted graph, the closer $K$ is to $1$, the closer the graph is to being a complete graph. Conversely, the closer $K$ is to $0$, the closer the graph is to being a set of disconnected vertices.
 
For a graph clustered into $\ell$ clusters, we also define the mean intra-cluster connections ratio, mean intra cluster density, as 
\begin{align*}
& \bar{K}_{\mbox{intra}} = \frac{1}{\ell} \sum_{i =1}^\ell \kappa_i = \frac{1}{\ell} \sum_{i=1}^{\ell} \frac{ \vert E_{ii} \vert }{ 0.5 \times n_i (n_i-1) } \\
& \text{where, } \\
& \kappa_i = \left\{ \begin{array}{c c} \frac{ \vert E_{ii} \vert }{ 0.5 \times n_i (n_i-1) } & \text{if } n_i \geq 2 \\
 0 & \text{otherwise}  \end{array} \right.  \; .
 \end{align*}
The mean intra-cluster connections ratio is the mean ratio of the number of edges within each cluster over the maximum number of edges that could possibly connect the vertices within each cluster. Each term in the summation, denoted as $\kappa_i$, represents each cluster's internal density, the density of the induced subgraph formed by its vertices and the edges connecting them. It is a measure of how closely each cluster is to being a clique. In the unweighted case, each $\kappa_i$ always lies on the interval $[0,1]$, with a value of $0$ indicating a cluster is just a set of disconnected vertices and a value of $1$ indicating that a cluster is a clique. At the aggregate level, $\bar{K}_{\mbox{intra}} $ is the sample mean of the individual terms $\kappa_i$ and also lies in the interval $[0,1]$. Values close to $0$ indicate poorly connected clusters on average, while values closer to $1$ indicate densely connected clusters on average.

Finally, we define the mean inter-cluster connections ratio, mean inter-cluster density, as 
\begin{eqnarray*}
\bar{K}_{\mbox{inter}} &=& \frac{1}{0.5 \times \ell(\ell - 1)}  \sum_{i=1}^{\ell} \sum_{j=i+1}^{\ell} \kappa_{ij} \\
&=&  \frac{1}{0.5 \times \ell(\ell - 1)} 
 \sum_{i=1}^{\ell} \sum_{j=i+1}^{\ell} \frac{ \vert E_{ij} \vert }{ 0.5 \times \left( (n_i + n_j)(n_i + n_j - 1) - n_i(n_i -1) - n_j(n_j -1) \right) } \\
 &=&  \frac{1}{0.5 \times \ell(\ell - 1)}  \sum_{i=1}^{\ell} \sum_{j=i+1}^{\ell} \frac{ \vert E_{ij} \vert }{ n_i \times n_j  } \; .
 \end{eqnarray*}
 
The mean inter-cluster connections ratio is the mean ratio of the number of edges joining vertices in a pair of clusters $(c_i,c_j)$ where $c_i$ and $c_j$ are distinct clusters, over the total number of edges that could possibly connect each pair of vertices across the cluster pair $(c_i,c_j)$. Each term in the double summation is the density of the induced bipartite graph formed by the vertices in each cluster pair, when we ignore the edges that join vertices within each cluster and only consider edges between vertices of either clusters of the pair. It is a measure of how close two clusters $i$ and $j$  are from a biclique, when considering only edges that have endpoints in either cluster. Here again, in the unweighted case these terms also lie in the interval $[0,1]$. A value of $0$ indicates no connection between a pair of clusters and a value of $1$ indicates the pair of clusters forms a biclique, when we ignore the intra-cluster edges. At the aggregate level, $\bar{K}_{\mbox{inter}} $ is the sample mean of the individual $\kappa_{ij}$ and also lies in the interval $[0,1]$. Values close to $0$ indicate low inter-cluster connections, on average, a desirable feature indicating strong cluster partitions. On the other hand, values closer to $1$ indicate improperly partitioned clusters, on average.

\subsubsection{Limitations:} 
Our inter- and intra-cluster measures are means and are therefore sensitive to outliers. For example, a few very dense clusters in an otherwise poorly clustered graph or very sparse clusters in an otherwise well clustered graph may render intra-cluster density uninformative. Mean inter-cluster density can also be affected in a similar way by outliers. A small number of densely inter-connected cluster pairs in an otherwise well clustered graph or a small number of sparsely inter-connected ones in an otherwise poorly clustered graph may also render mean inter-cluster density uninformative. 

Our inter- and intra-cluster measures are also unweighted means, which implies they are unaffected by cluster or cluster pair sizes. This immunity to cluster sizes is a deliberate design feature. It makes our measures immune to resolution-limit, a common shortcoming of clustering quality functions. A more detailed discussion of this phenomenon can be found in Section~\ref{axioms}. 

Nevertheless, precautions should be taken with conclusions in cases where cluster or cluster pair sizes are very heterogenous. For example, an algorithm may cluster a few small clusters very well and lump the remaining vertices into one or a few large sparsely connected clusters, which would result in a high mean intra-cluster density, even if the graph is arguably poorly clustered. 

Fortunately, in most cases, we expect that an inaccurately inflated mean intra-cluster density will also result in a high mean inter-cluster density. Likewise, we expect that an inaccurately low mean inter-cluster density  will also result in a low mean intra-cluster density. However, as with any other analysis based on means, the sample should also be carefully examined. In this specific case, particular attention should be paid to individual cluster and cluster pair sizes and to the individual $\kappa_i, \kappa_{ij}$ being averaged.

It should also be mentioned that in cases where the connectivity patterns of the clusters are very noisy (high variance), the median of the inter-/intra-cluster densities can be used in lieu of the mean. This substitution can produce more robust measures of inter- and intra-cluster connectivity. Unfortunately, it also makes interpretation and significance testing less obvious. 

\subsubsection{Illustrative Example:}
\begin{figure}[]
\centering
\includegraphics[height=65mm]{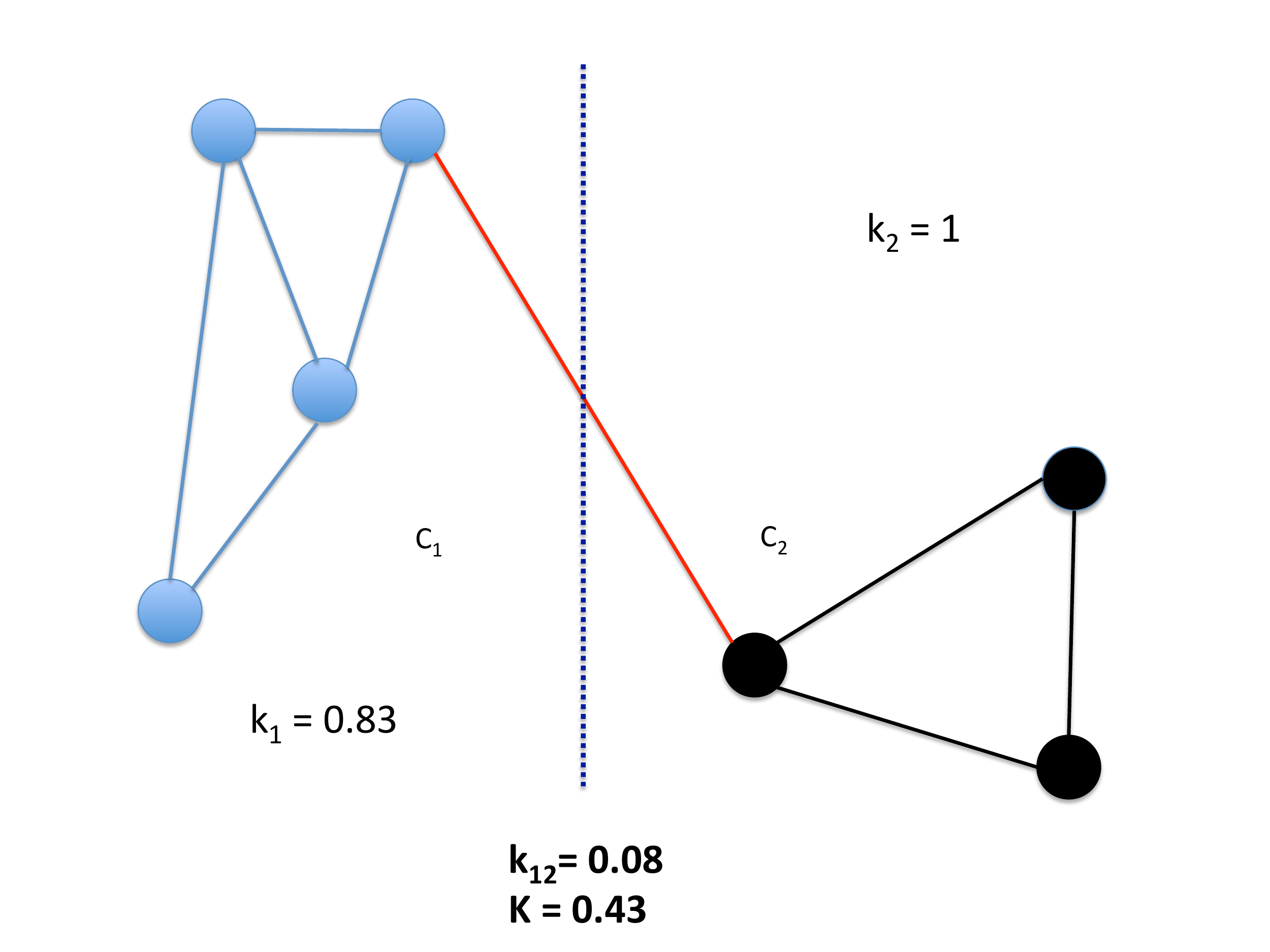}
\caption{Well-Clustered Graph} \label{densities}
\end{figure}

In the previous section, we define a well clustered graph as one where the inequalities $\bar{K}_{\mbox{inter}} < K < \bar{K}_{\mbox{intra}}$ hold. We illustrate this definition using Figure~\ref{densities}, which contains what is arguably a well labelled (clustered) graph with two clusters.  
If we compute the mean inter- and intra-cluster densities and compare them to the graph's global density, we see the inequalities described in Section~\ref{kappas} hold:
\begin{eqnarray*}
\bar{K}_{\mbox{intra}} &=& \frac{1}{2} (1 + 0.83) = 0.92 \\
\bar{K}_{\mbox{inter}} &=& \left( \frac{1}{0.5 \times 2 \times 1} \right) \frac{1}{4 \times 3} = 0.08 \\
K &=& \frac{9}{0.5 \times 7 \times 6} = 0.43 \\
\\
& \Rightarrow & \bar{K}_{\mbox{inter}} < K < \bar{K}_{\mbox{intra}} 
\end{eqnarray*}

\subsection{Statistical Interpretation of the Kappas and Clustering Quality} \label{interpret}
One of the appealing features of our Kappas lies in their statistical definition. The statistical definition provides a means to formally interpret them, assess their significance and compare their differences. Such comparisons and tests are problematic with many currently used quality measures. For example, in the case of modularity, these problems were identified by both Fortunato \cite{FortunatoLong2010} and Traag et al. \cite{Signif13}.

In the unweighted graph case, densities can be modeled and interpreted as Bernoulli probabilities. Under this Bernoulli model, the number of trials is the maximum number of possible edges, while the number of successes is the actual number of edges. The empirical estimate of the unconditional probability of connection between any two vertices on the graph is given by the graph's global density, $K$. Meanwhile, the conditional probability of a connection, given both nodes are in cluster i, is given by its internal density, $\kappa_i$. The conditional probability of connection, given one node is in cluster i and the other in cluster j, is given by their inter-cluster density, $\kappa_{ij}$. In each case, the quantities $K$, $\kappa_i$ and $\kappa_{ij}$ are the ratios of the observed number of edges over the maximum possible number of edges.

The cluster and cluster pair conditional probabilities of connection can also be generalized to obtain graph level values. This generalization provides probabilities of connection for all pairs of vertices within clusters or across cluster pairs. To obtain graph-wide estimates of these conditional probabilities, we take their means over all clusters or cluster pairs. Their means, $\bar{K}_{\mbox{inter}}$ and $\bar{K}_{\mbox{intra}}$ correspond to the graph level Bernoulli probabilities of two vertices sharing an edge, within clusters or across cluster pairs. 

In the weighted case, these empirical estimates of connection probability become the mean edge weights. The quantities $K$, $\bar{K}_{\mbox{intra}}$  and $\bar{K}_{\mbox{inter}}$ are the empirical estimate of edge weights connecting two nodes anywhere on the graph, within or between clusters. 

At this point, it must be mentioned that in estimating edge probabilities or weight means, our model assumes all clusters are equally likely. We do not weight our estimates by cluster  or cluster pair size deliberately, to avoid resolution-limit. As mentioned earlier, a more detailed discussion of resolution-limit is provided in Section~\ref{axioms}.
 
By exploiting the statistical interpretation of our clustering measures, we can offer a statistical description of good clustering. We posit that a good clustering will partition the graph such that the probability there exists an edge ($e_{ij}$) between two arbitrary nodes $i$ and $j$ is lower than the probability a connection exists if these nodes are in the same cluster (i.e., if $c_i = c_j$) and higher than when they belong to different clusters (i.e., $c_i \neq c_j$). Mathematically, we expect the following inequalities  to hold in cases where the clustering returned by an algorithm is of good quality ($\hat{P}$ denotes the empirical estimate of the probablities):
\[
\hat{P}\left[ e_{ij} \vert c_i \neq c_j \right] < \hat{P}[e_{ij}] < \hat{P}[e_{ij} \vert c_i = c_j ] \,.
\] 
In the case of a weighted graph, these empirical probability estimates become empirical estimates of the expected values of edge weights between arbitrary vertices, vertices within and vertices in different clusters. Consequently, we expect the following inequalities to hold ($\hat{E}$ denotes the empirical estimate of the expected values):
\[
\hat{E}[e_{ij} \vert c_i \neq c_j ] < \hat{E}[e_{ij}] < \hat{E}[e_{ij} \vert c_i = c_j ] \, .
\]

To better illustrate the intuition behind them, we use our statistical interpretation to examine our inequalities in the context of the following three base-case examples: an algorithm which lumps all vertices into one single cluster, an algorithm which assigns each vertex to its own cluster and an algorithm which randomly assigns vertices to $k$ clusters. Together, these examples form the set of the worst possible degenerate clusterings.

\subsubsection{All Vertices Into One Single Cluster}
Of all three poor clustering base-cases, this case is the only one where intra-cluster density is greater than inter-cluster density. Because there is only one cluster, we have the following equalities and inequality:
\begin{align*}
 & \bar{K}_{\mbox{inter}} = P(e_{ij} \vert c_i \ne c_j) = 0 \\
& \bar{K}_{\mbox{intra}} = P(e_{ij} \vert c_i = c_j) = P(e_{ij}) = K  \\
& \Rightarrow \bar{K}_{\mbox{intra}} \ngtr K > \bar{K}_{\mbox{inter}} \, .
\end{align*}
In this case, inter-cluster density is equal to zero, by definition. As a result, intra-cluster density is automatically greater than inter-cluster density $(\bar{K}_{\mbox{intra}} > \bar{K}_{\mbox{inter}})$, if there is at least one edge. Nevertheless, intra-cluster density is not higher than the global density  $(\bar{K}_{\mbox{intra}}  \ngtr K)$. Our inequalities test does indeed detect the degenerate clustering, since the first necessary condition is not met.

Generally, lumping all vertices into one cluster is a symptom of a poor clustering algorithm. However, there are instances where a graph may not be clusterable. It may not be formed by clusters, sets of dense subgraphs contained within a wider sparser graph. In such cases, assigning all its vertices to the same cluster is arguably the best possible clustering. For example, the best clustering for the vertices of a complete graph is to assign the same label to all of them. 

The single-cluster clustering is a limiting case for our characterization of good-clustering, because it may inaccurately classify proper clusterings as being of bad quality. Fortunately, however, it is a rare occurrence. Nevertheless, the single-cluster clustering must be treated as a special case and clustering quality must be assessed by other means. In particular, the graph's global density, which is equal to intra-cluster density, in this case, remains a valid measure of clustering quality. Global density provides a measure to trivially judge the quality of clustering in the single-cluster case. Arguably, a graph whose vertices are all clustered into one cluster is well clustered only if its global density is very high. It is poorly clustered otherwise.

\subsubsection{Single-vertex Clusters}
A clustering where each vertex is assigned to its own cluster is another base-case degeneracy (in cases where the graph has at least one edge). While it can trivially be identified by a simple examination  of the clustering algorithm's labeling, our first necessary condition does not hold. The poor-quality clustering is correctly identified and the following relations hold:
\[
\bar{K}_{\mbox{intra}} = 0 < K = \bar{K}_{\mbox{inter}} \, .
\]

\subsubsection{Random Cluster Assignment}
This degenerate clustering is not easily detectable by a simple examination of clustering results, unlike the previous two instances. In fact, when the number of vertices is large, this degeneracy is impossible to identify trivially. Fortunately, however, this is a case where our first necessary condition fails. The poor clustering is identified by these equalities:
\begin{align*}
& \bar{K}_{\mbox{inter}} = P(e_{ij} \vert c_i \ne c_j) = P(e_{ij}) = K \\
& \bar{K}_{\mbox{intra}} = P(e_{ij} \vert c_i = c_j) = P(e_{ij}) = K  \\
& \Rightarrow \bar{K}_{\mbox{intra}} = K = \bar{K}_{\mbox{inter}} \, .
\end{align*}
Because it is not possible to trivially identify this degeneracy, we consider it as the canonical example of a bad clustering.  For this reason, we use the random cluster labeling as our null model and as a null hypothesis for our statistical tests, which are described in the next section.

\subsection{Link to K-means}
K-means is one of the most widely used clustering algorithms. It applies to situations where covariates are quantitative and the Euclidian distance separating them is known \cite{ESL09}. Squared Euclidian distances are then used as dissimilarity measures between covariates. K-means is an iterative algorithm that groups covariates into clusters such that intra-cluster dissimilarity (distance) is minimized. Resulting clusters are formed by grouping covariates according to their similarity.

In the case of graphs, we typically do not have the benefit of inter vertex distances. We use density of connections as a measure of similarity. Under our definition of clustering quality, clusters are formed by vertices that are more similar to each other than to vertices in other clusters. Arguably, our definition corresponds to the rationale behind the K-means algorithm. An equivalent average-case relationship between intra cluster, inter cluster and global distances between covariates is expected to hold in instances of successful K-means clusterings. In such cases, it is naturally expected that the mean distances between covariates within clusters will be smaller than between covariates belonging to different clusters. In the case of a successful K-means clustering, we expect the following inequalities to hold:
\[
\bar{D}_{\mbox{intra}} < \bar{D} < \bar{D}_{\mbox{inter}} \, .
\]
In these inequalities, $\bar{D}_{\mbox{intra}}, \bar{D}, \bar{D}_{\mbox{inter}}$ denote the mean Euclidian distances between covariates within the same cluster, between any two covariates regardless of their cluster and between covariates belonging to different clusters. We note that in the case of K-means the inequalities are reversed, since the quantities being compared are dissimilarities.

\subsection{A Complete Clustering Quality Assessment Routine} \label{complete}
Our quality assessment routine consists of two steps, the verification of each of our two necessary conditions whose combined fulfillment forms a sufficient (average case) condition. In order to conclude on the quality of a clustering, we must verify that our first set of inequalities holds. In the event they do, we must also ensure they hold at a statistically significant level.

The first step consists of the computation of global, inter- and intra-cluster densities and their comparison through our first necessary condition inequalities test. In this first step, we verify that the inequalities $\bar{K}_{\mbox{inter}} < K <\bar{K}_{\mbox{intra}}$ hold. We verify that mean intra-cluster density is higher than global density and that global density is also higher than mean inter-cluster density. These inequalities form the foundations of our statistical (average case) definition of good clustering.

Once it has been established that our first set of necessary conditions have been met, we test the statistical significance of the inequality $\bar{K}_{\mbox{intra}} > \bar{K}_{\mbox{inter}} \Leftrightarrow \gamma = \bar{K}_{\mbox{intra}} - \bar{K}_{\mbox{inter}} > 0 $, sample size permitting. This second step consists of a statistical significance test of the difference between intra- and inter-cluster densities, which we call $\gamma$. A detailed description of statistical significance tests is provided in the next section.

\begin{itemize}
\item \textbf{Clustering Quality Assessment Routine:}
\begin{itemize}
\item Use clustering algorithm labels to compute the Kappas
\item Numerically verify that the inequalities $\bar{K}_{\mbox{inter}} < K < \bar{K}_{\mbox{intra}}$  hold (first necessary condition) 
\item If they don't and if the number of clusters is greater than one, conclude the algorithm has poorly clustered the graph
\item If they don't and if the number of clusters is one, use global density to assess clustering quality
\item If they do hold and if the number of clusters is sufficiently large, perform statistical test to verify significance (second necessary condition)  of the difference \[ \gamma = \bar{K}_{\mbox{intra}} - \bar{K}_{\mbox{inter}} \] 
\item If testing more than one algorithm that meet all benchmarks above, compare p-values to find best algorithm
\end{itemize}
\end{itemize}
In closing, we note that in many cases a simple examination of the inequalities in the first necessary condition, along with some domain expertise may be sufficient to draw conclusions on clustering quality. 

\section{Hypothesis Testing} \label{stattest}
As mentioned earlier, because our measures of clustering are also graph statistics, we can push our analysis further and verify our second necessary condition, which is described in Section~\ref{qualitydefn}. Through this test, we ensure the difference between intra- and inter-cluster densities is statistically significant. Of course, we could also test the significance of the difference between intra-cluster and global densities and between global density and intra-cluster density. However, these multiple tests would be redundant and would complicate the testing process. After all, according to all definitions of good clustering, the benchmark is that intra-cluster connectivity (density) be greater than inter-cluster connectivity (density). 

The stochasticity which makes hypothesis testing possible stems from the fact each clustering is one sample of an unobserved  distribution of vertex labelings. Each graph's clustering, as returned by one particular clustering algorithm, can be understood as being one sample drawn from this unobserved distribution of all possible clusterings into the same number of clusters as those identified by the algorithm (or set by external parameter) for the graph under study. The quantities, $\bar{K}_{\mbox{intra}} \text{ and } \bar{K}_{\mbox{inter}}$ are the corresponding sample means of intra- and inter-cluster densities. 

To formally confirm statistical significance, we use a modified version of the standard Student's t-test, which is described in Section~\ref{test stat}. It must be emphasized that because our performance measures are sample estimates of a mean, we do not face the problem of assigning them a distribution. This clear statistical definition is in contrast to the difficulty of assigning a distribution to modularity. Such difficulty renders formal statistical tests of its significance non-informative, as highlighted by Fortunato in 2010 \cite{FortunatoLong2010}.  Our Kappas are assumed to be distributed about their true value according to a Gaussian distribution, on the basis of the Central Limit Theorem, when sample size (number of clusters) is sufficiently large. Our computational experiments also reveal they remain Gaussian even for smaller sample sizes.

Here, we streamline our statistical test. In our previous article \cite{PMEtAlWAW18}, we conducted two separate tests.  We formulated two null hypotheses, $\bar{K}_{\mbox{intra}} = K$ and $\bar{K}_{\mbox{inter}} = K$, to avoid the effects of a possible correlation between $\bar{K}_{\mbox{intra}}$ and $\bar{K}_{\mbox{inter}}$ ($K$ is a graph constant, not the result of a clustering). However, since our ultimate goal is to formally compare intra- and inter-cluster densities, we adapt the standard t-test to overcome any possible correlation and allow for a direct comparison of these graph statistics. This new test verifies that a clustering meets the second necessary condition for being classified as a good clustering. If it does meet this condition, we expect the inequalities $\bar{K}_{\mbox{intra}} > \bar{K}_{\mbox{inter}}$ to hold at a reasonable significance level (e.g., $\alpha = 0.95$). 

Our test can also be used when comparing two or more algorithms' performances on a given graph. In such cases, in order to conclude algorithm `a' is better than algorithms `b', 'c',$(\dots)$, we should observe better (smaller) p-values, $p_a < p_b < p_c < (\ldots)$. Although this procedure is not a formal statistical test, it is a valid and easily applicable heuristic.

Finally, let us note that our statistical definition also allows for uncertainty in the connectivity data, another open problem which was identified by Holder et al. in 2016 \cite{futureChallenges2016}. Unlike modularity and conductance, our measures are defined as statistical measurements with associated standard errors, not deterministic quantities.

\subsection{Null Hypothesis} \label{nulls}
Under the null hypothesis, the algorithm is assumed to offer a random assignment of nodes to clusters. When nodes are randomly assigned cluster labels, we expect no significant difference between $\bar{K}_{\mbox{intra}}$  and $\bar{K}_{\mbox{inter}}$, as described in Section~\ref{interpret}.

Here, we note that our null hypothesis does not rely on any generative model for the graph, unlike modularity which uses the configuration model as a null model. In fact, in our approach, the graph is not random, but is fixed.  Instead, under the null hypothesis, the clusters are random.

Our significance test is an assessment of the statistical significance of the quality of a clustering returned by an algorithm. We argue that in the case of a good clustering, the gap between intra- and inter-cluster densities should be statistically significant (test 1). We also use the p-values of this test to heuristically compare the quality of the clusterings of a specific graph returned by two or more algorithms, as mentioned earlier (test 2).

\begin{align*}
& \text{\bf Test 1} & \\
&H_0: \quad \bar{K}_{\mbox{intra}} = \bar{K}_{\mbox{inter}} \Leftrightarrow \gamma = \bar{K}_{\mbox{intra}} - \bar{K}_{\mbox{inter}} = 0 \\
&H_a: \quad \bar{K}_{\mbox{intra}} > \bar{K}_{\mbox{inter}} \Leftrightarrow \gamma = \bar{K}_{\mbox{intra}} - \bar{K}_{\mbox{inter}} > 0 \\
\\
& \text{\bf Test 2} \, \text{(heuristic, two algorithm comparistion)} & \\
&H_0: \quad \bar{K}_{\mbox{intra}}^{(1)} = \bar{K}_{\mbox{inter}}^{(1)} = \quad \bar{K}_{\mbox{intra}}^{(2)} = \bar{K}_{\mbox{inter}}^{(2)} \Leftrightarrow \gamma_1 = \gamma_2 = 0\\
&H_a: \quad \gamma_1 > \gamma_2 
\end{align*}

\subsection{Test Statistics} \label{test stat}
To test the hypotheses in the previous section, we compute the following test statistics, one for assessing the significance of $\gamma$ (stat test 1) and one for comparing the quality of the clusterings as labeled by two or more clustering algorithms (stat test 2). 
\begin{itemize}
\item Stat Test 1: \[ t = \frac{\gamma}{s.e.} \]
\item Stat Test 2 (heuristic): 
\begin{itemize}
\item Compute $t_a$ and $t_b$, the significance tests (test 1, here above) for algorithms `$a$' and `$b$' on a given graph
\item Obtain the respective p-values, $p_a$ and $p_b$
\item If $p_a - p_b > 0$, conclude algorithm `$b$' returned a better, more statistically significant, clustering
\item Note this is a heuristic decision tool, not a formal statistical test
\end{itemize} 
\end{itemize} 

The modification we make to the t-test lies in the computation of the standard error ($s.e.$) and in the degrees of freedom of the t statistic. In the classic t-test, the t statistic for Test 1 would be computed as follows.
\begin{eqnarray*}
t &=& \frac{\gamma}{s_{\gamma}} \\
s_{\gamma} &=& \sqrt{ \frac{s^2_{\mbox{intra}} }{\vert C \vert} + \frac{s^2_{\mbox{inter}}}{ 0.5 \times \vert C\vert \times (\vert C\vert -1)} } \\
s^2_{\mbox{intra}} &=& \frac{1}{ \vert C \vert - 1 } \sum_{i=1}^{\vert C \vert} \left( \kappa_i - \bar{K}_{\mbox{intra}} \right)^2 \\
s^2_{\mbox{inter}} &=& \frac{1}{ 0.5 \times \vert C\vert \times (\vert C\vert -1) - 1 } \sum_{i=1}^{\vert C \vert} \sum_{j = i+1}^{\vert C \vert} \left( \kappa_{ij} - \bar{K}_{\mbox{inter}} \right)^2
\end{eqnarray*}

Instead, we use Monte-Carlo simulation to compute the t-statistic directly. The steps in this computation are described below:
\begin{enumerate}
\item Randomly label nodes as belonging to one of the $\vert C \vert$ clusters identified by the algorithm (null hypothesis)
\item Compute $\gamma = \bar{K}_{\mbox{intra}} - \bar{K}_{\mbox{inter}}$, under the null hypothesis
\item Repeat $r > 30$ times and compute the variance $Var(\gamma)$
\item Use $s.e. = \sqrt{Var(\gamma)}$
\item Degrees of freedom for the test are given by the number of simulation runs, $r$, minus one, $d.f. = (r-1)$
\end{enumerate}

In addition to sidestepping the issue of possible dependencies between $\bar{K}_{\mbox{intra}}$ and $\bar{K}_{\mbox{inter}}$, the main feature of our modified t-test is that it remains computable even in cases where the standard error cannot be estimated by applying the usual scaling to the variance of the results. For example, in cases where the graph is disconnected, the variance $s^2$ (and standard error) of $\bar{K}_{\mbox{inter}}$ may be hard, even impossible, to estimate accurately. By using Monte-Carlo simulation to estimate standard error, we are able to overcome the obstacles posed by such situations.

Of course, we could also apply the Monte-Carlo method to obtain p-values directly, without using t-statistics and the Student distribution. Proceeding in this way would not only circumvent any possible dependencies but would also be independent of any distributional assumptions on the null hypothesis. Unfortunately, such a procedure would also be very computationally demanding and could render testing infeasible. Fortunately, as we show in the next section, the Gaussian distribution offers a very good model of our null hypothesis, even when sample sizes (number of clusters) are small $(N < 30)$.

\subsection{Empirical Examination of the Null Distribution}
Under our null hypothesis, our canonical example of poor clustering, the difference $\gamma = \bar{K}_{\mbox{intra}} - \bar{K}_{\mbox{inter}}$ has an expected value of zero. It is also approximately Gaussian. Indeed, under the null, cluster labels are assigned to vertices randomly. In this case, the cluster label of a vertex is independent of its connections to other vertices. The symmetry of the distribution of $\gamma$ stems from the fact it is a difference of two sample means. 

\begin{figure} 
\centering
\subfloat[ER, 35 runs, 12 clust]{ \includegraphics[width = 0.3\textwidth]{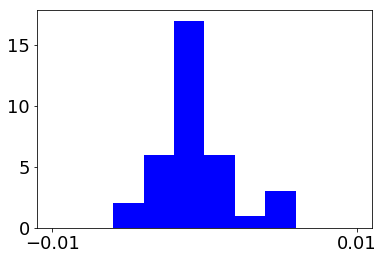} }
\subfloat[ER, 100 runs, 12 clust]{ \includegraphics[width = 0.3\textwidth]{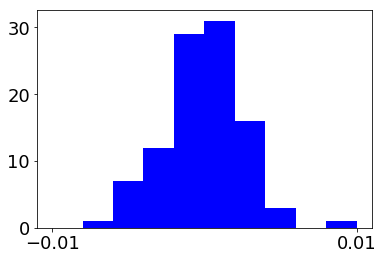} } 
\subfloat[ER, 1000 runs, 12 clust]{ \includegraphics[width = 0.3\textwidth]{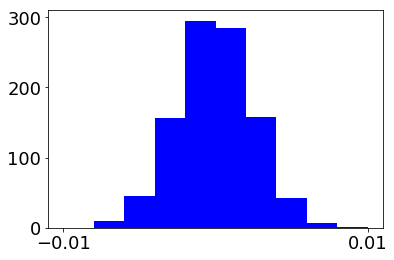} }
\\
\subfloat[ER, 35 runs, 24 clust]{ \includegraphics[width = 0.3\textwidth]{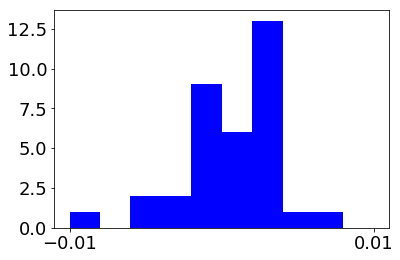} }
\subfloat[ER, 100 runs, 24 clust]{ \includegraphics[width = 0.3\textwidth]{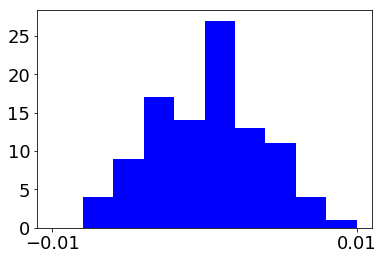} } 
\subfloat[ER, 1000 runs, 24 clust]{ \includegraphics[width = 0.3\textwidth]{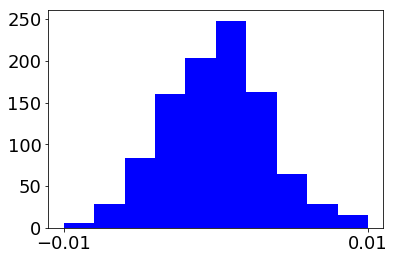} }
\\
\subfloat[CC, 35 runs, 12 clust]{ \includegraphics[width = 0.3\textwidth]{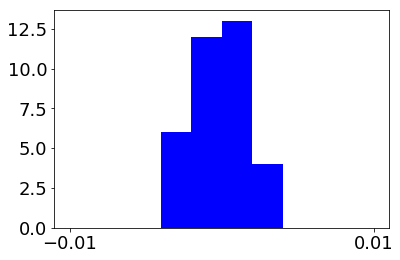} }
\subfloat[CC, 100 runs, 12 clust]{ \includegraphics[width = 0.3\textwidth]{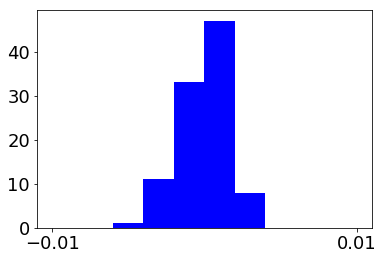} } 
\subfloat[CC, 1000 runs, 12 clust]{ \includegraphics[width = 0.3\textwidth]{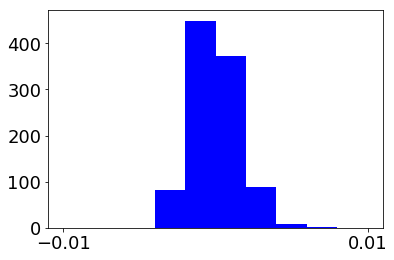} }
\\
\subfloat[CC, 35 runs, 24 clust]{ \includegraphics[width = 0.3\textwidth]{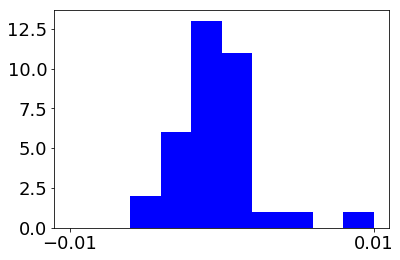} }
\subfloat[CC, 100 runs, 24 clust]{ \includegraphics[width = 0.3\textwidth]{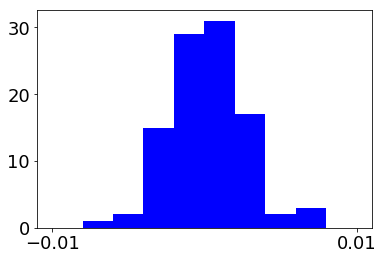} } 
\subfloat[CC, 1000 runs, 24 clust]{ \includegraphics[width = 0.3\textwidth]{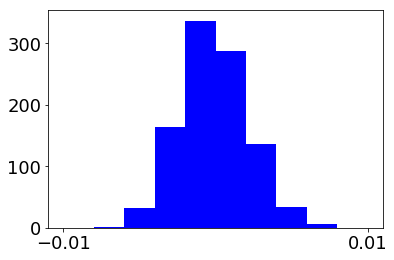} }
\caption{Null Distributions of $\gamma$, Number of Runs and Number of Clusters}
\label{hists}
\end{figure}

\begin{figure} 
\centering
\subfloat[ER, 35 runs, 12 clust]{ \includegraphics[width = 0.3\textwidth]{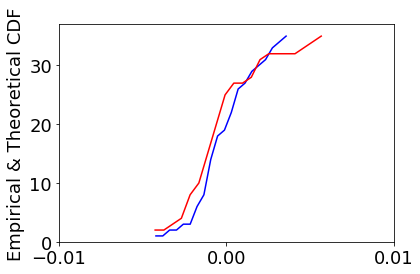} }
\subfloat[ER, 100 runs, 12 clust]{ \includegraphics[width = 0.3\textwidth]{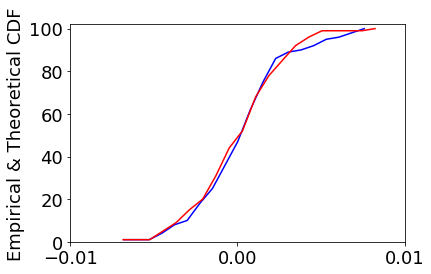} } 
\subfloat[ER, 1000 runs, 12 clust]{ \includegraphics[width = 0.3\textwidth]{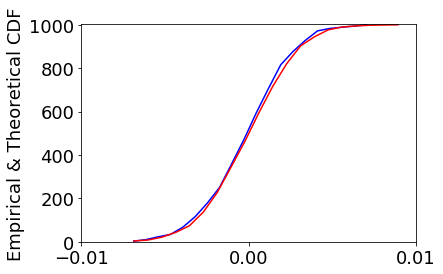} }
\\
\subfloat[ER, 35 runs, 24 clust]{ \includegraphics[width = 0.3\textwidth]{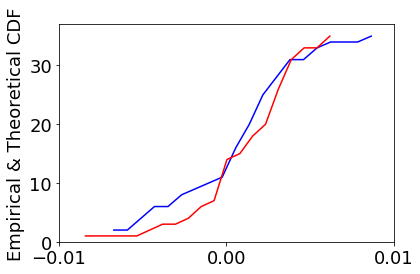} }
\subfloat[ER, 100 runs, 24 clust]{ \includegraphics[width = 0.3\textwidth]{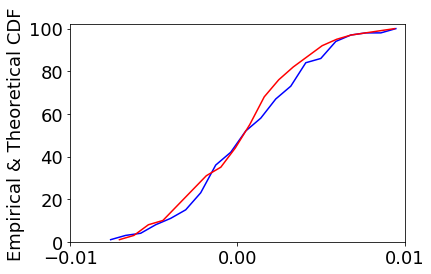} } 
\subfloat[ER, 1000 runs, 24 clust]{ \includegraphics[width = 0.3\textwidth]{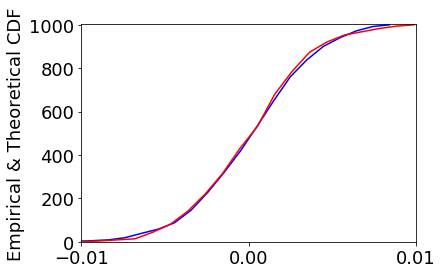} }
\\
\subfloat[CC, 35 runs, 12 clust]{ \includegraphics[width = 0.3\textwidth]{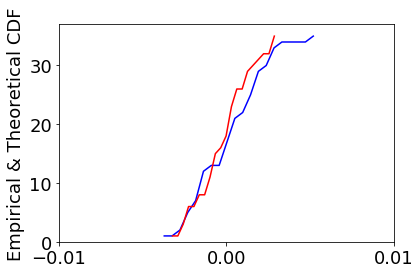} }
\subfloat[CC, 100 runs, 12 clust]{ \includegraphics[width = 0.3\textwidth]{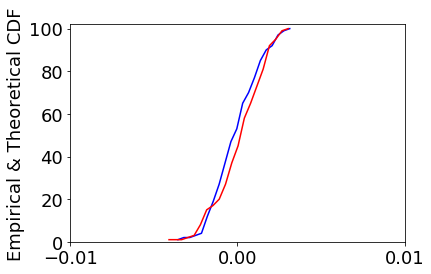} } 
\subfloat[CC, 1000 runs, 12 clust]{ \includegraphics[width = 0.3\textwidth]{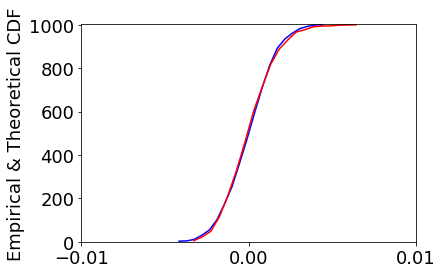} }
\\
\subfloat[CC, 35 runs, 24 clust]{ \includegraphics[width = 0.3\textwidth]{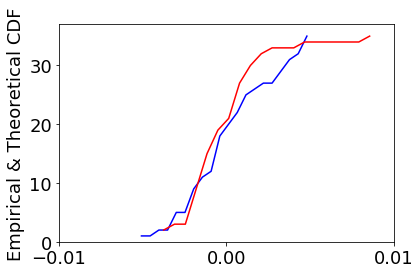} }
\subfloat[CC, 100 runs, 24 clust]{ \includegraphics[width = 0.3\textwidth]{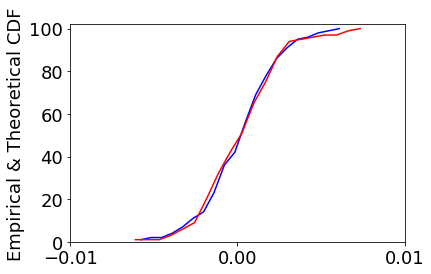} } 
\subfloat[CC, 1000 runs, 24 clust]{ \includegraphics[width = 0.3\textwidth]{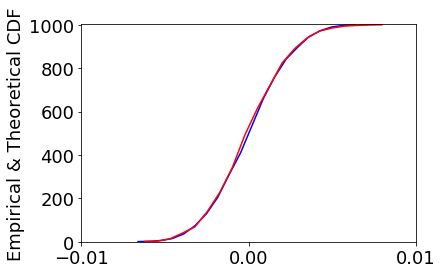} }
\caption{Null Distributions of $\gamma$ in Red, Number of Runs and Number of Clusters, Gaussian with Same Mean and Std Dev in Blue}
\label{cdfs}
\end{figure}

To empirically validate our statements about the distribution of $\gamma$ under the null, we simulate random node labelings on two different synthetic graphs. The first graph is an Erd\H{o}s-R\'enyi (ER) graph of $1,000$ vertices and edge probability of $\frac{1}{3}$. The second is a connected caveman (CC) graph of $10$ (quasi-)cliques of $100$ vertices. Each (quasi-)clique has one edge re-assigned so it connects to one vertex in another cluster. These graphs were chosen, because they lie at either end of the spectrum of structured-unstructured graphs. We then simulate a random vertex labeling of $12$ and $24$ clusters, repeat $r$ times and compute the means and standard deviations of $\gamma$. To complete our comparisons, we examine the sample statistics of our simulated data and its empirical distribution. We also superimpose its percentiles over the percentiles of a Gaussian distribution with the same mean and standard deviation. The results are shown in Table~\ref{simres}, Figures~\ref{hists} and~\ref{cdfs}.

\begin{table}[]
\centering
\caption {Mean and Std Dev of $\gamma$ Under the Null} \label{simres} 
\begin{tabular}{|c|| c| c| c| c|}
\hline
Graph & Num Runs & Num Clusters & Mean gamma & Std gamma \\
\hline
ER1000 & 35   & 12 & -0.0005 & 0.0023 \\
ER1000 & 100  & 12 & -0.0001 & 0.0026 \\
ER1000 & 1000 & 12 & 0.0000  & 0.0025 \\
ER1000 & 35   & 24 & 0.0010  & 0.0030 \\
ER1000 & 100  & 24 & 0.0001  & 0.0035 \\
ER1000 & 1000 & 24 & 0.0001  & 0.0034 \\
CC1000 & 35   & 12 & -0.0002 & 0.0016 \\
CC1000 & 100  & 12 & 0.0001  & 0.0015 \\
CC1000 & 1000 & 12 & 0.0000  & 0.0015 \\
CC1000 & 35   & 24 & -0.0004 & 0.0024 \\
CC1000 & 100  & 24 & 0.0001  & 0.0023 \\
CC1000 & 1000 & 24 & -0.0001 & 0.0022 \\
\hline
{\bf Mean} &          &              & { \bf 0.0000}    & { \bf 0.0024}  \\
\hline
\end{tabular}
\end{table}

The histograms in Figure~\ref{hists} reveal that $\gamma$ is roughly symmetrically distributed about its mean of zero, even with a small number of runs. That symmetry becomes more apparent as the number of runs increases. The numerical results in Table~\ref{simres} confirm its mean is always roughy equal to zero and standard deviation also roughly equal to zero. More importantly, our results also confirm the empirical percentiles of simulated $\gamma$ and those of a Gaussian with the same mean and standard deviation approximately coincide, especially when the number of runs is large. This coincidence can be seen in Figure~\ref{cdfs}, where the percentiles of $\gamma$ converge towards theoretical Gaussian percentiles, as the number of runs increases. It should be noted that cases where the histograms appear skewed are due to the relatively small number of runs with respect to the number of clusters. 

In summary, our node labelling simulations reveal that even a very small number of runs (e.g., $35$ runs) can offer a very accurate estimate of the standard deviation under the null, of the standard error of our modified t-statistic. They also confirm that under the null $\gamma$ has an expected value of zero. However, the most interesting result is that under the null $\gamma$ remains approximately Gaussian, even in instances where the cluster number is small (12 and 24) and well below the usual sample size required to invoke the Central Limit Theorem $(N \ge 30)$.

\subsection{Scalability of the Modified t-test} 
With the mass of large datasets that are now commonly studied, it is important to consider the scalability of any clustering performance measurement technique. Indeed, any test that cannot be applied to larger graphs is not suited to the emerging area of complex networks. However, because of the stability of the null and its Gaussian distribution, our performance variables and associated test statistics are indeed applicable to such data sets.

In particular, the stability of the standard distribution of the null eases the estimation of our t-statistic's standard error, since its computation doesn't require a large number of simulation runs. More importantly, it also allows us to use a modified t-statistic to estimate p-values, instead of resorting to costly direct empirical Monte-Carlo estimation. Additionally, it should be noted that computing $\bar{K}_{\mbox{intra}}$ and $\bar{K}_{\mbox{inter}}$ can be obtained by computing each cluster's internal density and each cluster pairs' common density separately. Breaking up computations in such a way allows for easy parallelization.

\section{Axioms for a Good Clustering Quality Function} \label{axioms}
In this section, we review the axioms that define a good clustering quality function and describe how our Kappas meet these axioms. Multiple authors have presented axioms defining a good clustering quality function \cite{Kleinberg03,AckerBD08,Laar14,Pitsoulis2018}. Although these publications use different terminology, their axioms share common features and are rooted in the seminal work of Kleinberg \cite{Kleinberg03}. 

We combine the recent work of Van Laarhoven and Marchiori \cite{Laar14} and the more recent work of Kehagias and Pitsoulis \cite{Pitsoulis2018} to draw a list of axioms defining the desirable properties of a clustering quality function for graphs. Our combined list of axioms contains ten axioms which characterize a good clustering quality function.

\subsection{Axioms of Van Laarhoven and Marchiori}
Most of the axioms that a good clustering quality function must meet are presented by Van Laarhoven and Marchiori \cite{Laar14} and are  listed below. In their work, these authors also prove modularity fails to meet two of these axioms, locality and monotonicity.
\begin{enumerate}
\item {\bf Permutation Invariance}: Clustering quality should not depend on specific labels. Instead, it should depend on global labeling structure. ({\it ``isomorphism invariance''} in Kehagias and Pitsoulis \cite{Pitsoulis2018})
\item {\bf Scale Invariance}: Clustering quality should remain unaffected when edge weights are scaled uniformly. Mathematically, this means that for a graph $G$, any clusterings $\left(C_i, C_j\right)$, constant $\alpha > 0$ and given quality function $Q(G,C)$, the following equivalence holds: \[Q(G,C_i) \leq Q(G,C_j) \Leftrightarrow Q(\alpha G,C_i) \leq Q(\alpha G,C_j) \, . \] 
While this axiom appears to apply mostly to weighted graphs, the idea behind it can be generalized to unweighted graphs. In the case of unweighted graphs, this axiom can be understood as meaning the equivalence above 
must hold when edge probabilities are uniformly scaled. 
\item {\bf Richness}: Given a finite set of vertices $V$ it is possible to re-arrange edges (or edge weights) to obtain an optimal partition.
\item {\bf Monotonicity and Consistent Improvement}: A ``consistent improvement'' means an increase in edge density (edge probability) within clusters or a decrease in inter-cluster edge density (edge probability) should not decrease the quality function. 
\item {\bf Locality}: 
\begin{enumerate}
\item Although Ackerman and Ben-David \cite{AckerBD08} have also defined locality in a very similar manner, we prefer the definition of Van Laarhoeven and Marchiori \cite{Laar14}. 
\item The latter is more flexible and does not rely on the assumption the number of clusters is known in advance. After all, when we are evaluating the performance of an algorithm, we do not want to impose a fixed number of clusters. First, in most cases, we don't know this number. Second, we also want to assess the validity of the number of clusters fed into or identified by the clustering algorithm under examination.
\item We use the definition that follows: ``(...) the contribution of a single cluster to the total quality should only depend on nodes in the neighborhood of that cluster (...) On the other hand, a quality function that is written as a sum over clusters, where each summand depends only on properties of nodes and edges in one cluster and not on global properties, is local.'' \cite{Laar14}.
\end{enumerate}
\item {\bf Continuity}: Here, the authors say ``A quality function Q is continuous if a small change in the graph leads to a small change in the quality.'' Essentially, this property ensures quality functions remain robust to noise in the graph structure.
\item {\bf Resolution-limit Free}: The partition remains optimal for any induced subgraph of the optimal partition. It was described as ``the limitations in detecting small community structures in a large network'', by McSweeney et al.  \cite{GameTheoretic14}. In essence, a resolution-limit free quality function's output is not disproportionately influenced by larger clusters to a point where the effect of smaller clusters is washed away and hidden.
\end{enumerate}

\subsection{Additional Axioms from Kehagias and Pitsoulis}
Many of the axioms of Kehagias and Pitsoulis \cite{Pitsoulis2018} are similar to those of Van Laarhoven and Marchiori \cite{Laar14}. In this section, we include some additional ones which are specific to the work of Kehagias and Pitsoulis. 
\begin{enumerate}
\setcounter{enumi}{6}
\item {\bf Perfectness}: ``(...) is based on the intuition that a union of disjoint complete graphs should exhibit perfect community structure''
\item {\bf Connectivity}: ``(...) is based on the intuition that a minimum requirement for a cluster to be classified as a community is that the associated induced subgraph should be connected''
\item {\bf Complementarity}: Let $Q$ be a uniformly scaled quality function on the interval $[0,1]$ and $G^c$ the complement graph of $G$, then if $Q$ is complementary, the following holds
\[
Q(G,C) = 1 - Q(G^c,C) \,.
\]
\end{enumerate}

\subsection{Axioms of Good Clustering Applied to the Kappas and Its Accompanying Tests}
In this section, we apply the axioms  of a good clustering quality function to our Kappas and their significance test. We describe how they meet all ten axioms we just listed. 
\begin{enumerate}
\item {\bf Permutation Invariance}: Cluster labels are only used to aggregate edge and vertex counts. A label permutation, swapping all node cluster labels between an arbitrary number of pairs of clusters, does not affect edge or vertex counts. Therefore, our estimates of intra- or inter-cluster density or the graph's global density remain unaffected.
\item {\bf Scale Invariance}: In the context of our tests, scale invariance must hold on two levels. Not only must the relative differences in the Kappas remain unaffected by the scaling, but the test statistics must also remain unaffected. A multiplication of edge probabilities or weights by a constant $\alpha > 0$ does not affect the relative differences in the Kappas, the t-test statistics and their degrees of freedom. Consequently, the conclusions of our tests are also unaffected. A full proof of this statement is trivial but lengthy. Essentially, it is based on the fact that our Kappas are non-negative numbers and the fact standard error scales linearly. Therefore, a multiplication by a non-zero positive constant does not affect the inequalities: \[
 \frac{a}{b} \lessgtr \frac{c}{d} \Leftrightarrow \frac{\alpha a}{\alpha b} \lessgtr \frac{\alpha c}{\alpha d}  \, . \]
 \item {\bf Richness}: This property means that the optimum is an achievable quantity. It is obvious that increasing intra-cluster density or decreasing inter-cluster density improves our test statistics' value and implies an improvement of clustering. In other words, the better the clustering, the greater the gap between global, intra- and inter-cluster density, the higher the t-statistic and the lower the p-value. For example, if we increase intra-cluster density and decrease inter-cluster density sufficiently, intra-cluster density will eventually reach its maximum value of one (case of unweighted graphs), while inter-cluster will reach its minimum value of zero (also unweighted graph case). Their difference, the quantity $\gamma$, will also reach its maximum value of one. The  significancel test statistics will also reach their optima (i.e., minimum p-value), if we modify densities sufficiently.
 
\item {\bf Monotonicity and Consistent Improvement}: This property is a consequence of the previous one.
\item {\bf Locality}: Our quality measurement quantities are means. They are sums over clusters or cluster pairs scaled by a constant. Each summand depends exclusively on its cluster or cluster pair. Our tests meet this property by definition: ``(..) a quality function that is written as a sum over clusters, where each summand depends only on properties of nodes and edges in one cluster and not on global properties, is local.'' \cite{Laar14}.
\item {\bf Continuity}: Continuity is a property of the mean, which provides a smoothed summary of a data set. A small improvement in clustering cannot result in a large fluctuation in the inter- or intra-cluster means, by construction. For example, a higher intra-cluster density within one or a few clusters cannot have a drastic effect on our graph-wide quantities. 
\item {\bf Perfectness}: The union of disjoint complete graphs would indeed have perfect scores. In such a case, each cluster's intra-cluster density would be equal to one and all inter-cluster densities would be qual to zero, by definition. \[
\bar{K}_{\mbox{intra}} = \frac{1}{\vert C \vert} \underbrace{\sum 1}_{\vert C \vert \times 1} = 1 > K > \bar{K}_{\mbox{inter}} = \frac{1}{0.5 \times \vert C \vert \left( \vert C \vert - 1 \right)} \sum 0 = 0 \]
Both necessary conditions would be met. The inequalities between global, intra- and inter-cluster densities would hold and $\gamma$ would reach its maximum value. The statistical test would classify the resulting $\gamma$ as significantly different from zero with a p-value approaching 0:
\[
t = \frac{(1-0)}{s.e.}, \, \text{ with } s.e \lll 0 
\]

\item {\bf Connectivity}: A set of disjoint vertices $v_i$ would have internal density \[
\kappa_{i} = 0 \leq K, \bar{K}_{\mbox{inter}}\] and would not meet our test for being a valid cluster. Such a set would trivially be classified as degenerate.
\item {\bf Complementarity}: Both our test statistics have number of edges in their numerators. 
\[
\kappa_i = \frac{ \vert E_{ii} \vert}{0.5 \times n_i (n_i-1)}, \; \kappa_{ij} = \frac{ \vert E_{ij} \vert}{n_i n_j}
\]

\begin{eqnarray*}
\kappa_{i^c} &=& \frac{ \vert E_{i^c i^c} \vert}{0.5 \times n_i (n_i-1)} \\
&=& \frac{0.5 \times n_i (n_i-1) - m_{ii}}{0.5 \times n_i (n_i-1)} \\
&=& 1 - \frac{m_{ii}}{ 0.5 \times n_i (n_i-1) } = 1 - \kappa_i   \\
\frac{1}{\vert C \vert} \sum_{i=1}^{\vert C\vert} \left(1 - \kappa_i \right) &=&  \frac{1}{\vert C \vert} \left( \vert C \vert - \sum_{i=1}^{\vert C\vert} \kappa_i \right)\\
&=& 1 - \frac{1}{ \vert C \vert } \sum_{i=1}^{\vert C\vert} \kappa_i  = 1 - \bar{K}_{\mbox{intra}} \; \square 
\end{eqnarray*}
Let $i^c, j^c $ be the complement of the bi-clique formed by clusters $i, j$, $ \vert E_{ij} \vert = m_{ij}$ be the number of edges with one end in each cluster, $n_i$ be the number of nodes in cluster $i$ and the total number of cluster-cluster pairs be denoted as $N = 0.5 \times \vert C \vert \left( \vert C \vert -1 \right)$.
\begin{eqnarray*}
\kappa_{i^c j^c} &=& \frac{\vert E_{i^c j^c} \vert}{n_i n_j} \\
&=& \frac{n_i n_j  - m_{ij}}{n_i n_j} = 1- \frac{m_{ij}}{n_i n_j} \\
\frac{1}{N} \left( \sum_{i,j=i+1}^N \left( 1- \kappa_{ij} \right) \right) &=& \frac{1}{N} \left( N - \sum_{i,j=1+1}^N \kappa_{ij} \right) \\
&=& 1 - \frac{1}{N} \sum_{i,j=1+1}^N \kappa_{ij} = 1 - \bar{K}_{\mbox{inter}} \; \square
\end{eqnarray*}
\item {\bf Resolution-limit Free}: If our first necessary condition (set of inequalities) is met, $\bar{K}_{\mbox{inter}} < K < \bar{K}_{\mbox{intra}}$, then we also expect $E(\kappa_{ij}) < K < E(\kappa_i)$ for all clusters $i$ and cluster pairs $i,j$. If our second necessary condition is met and we reject the null that $\bar{K}_{\mbox{intra}} = \bar{K}_{\mbox{inter}}$, then the corresponding null on subset statistics, $\bar{K}_{\mbox{intra}}^{(s)} = \bar{K}_{\mbox{inter}}^{(s)}$, will tend to be rejected as well. 

More concretely, it is important to note that neither $\bar{K}_{\mbox{intra}}$ nor $\bar{K}_{\mbox{inter}}$ are affected by individual cluster size relative to network size. They do not suffer from the resolution limit observed in modularity \cite{ResolLimitFortunato2007,Kehagias2013,MetricsScale16,guideFortunato16}. Very large (very small) clusters do not influence their values more than smaller (larger) ones, as with modularity. All terms in the sums are scaled by the total number of possible edges within each cluster or pair of clusters, which ensures they remain within the same order of magnitude regardless of size. In the case of unweighted graphs, they always lie on the $[0,1]$ interval. In the weighted case, they are always proportional to edge weight. This feature makes these measures robust to large ``mega-clusters'' that are often observed in real-world networks and to the problematic tendency of clustering algorithms to lump all vertices together in a few very large clusters \cite{guideFortunato16,ResolLimBrain17}. (Naturally, $K$ is a graph-wide measure that remains completely unaffected by cluster labelings and individual cluster sizes.)
\end{enumerate}

\section{Computational Experiments} \label{exps}
To empirically compare each competing performance measure's accuracy and responsiveness to various graph structures and cluster labelings, we subject them to a number of numerical stress test scenarios. We use simulated graphs and cluster labels. The full experimental set-up of our tests and scenario details are described in the next section. 

Overall, our goal is to test the accuracy and robustness of our clustering measures and compare their behavior to that of the two main clustering measures in the literature, modularity and conductance. Simulation is used to generate test scenarios where the clustering structure is known in advance and can be modified easily. These test scenarios are then used to examine and compare the sensitivities of the Kappas, modularity and conductance. Our scenarios include a number of contrived instances, but these are useful to stress test our metrics through extreme degenerate examples and compare their behavior to those of the more established measures. 

The overarching logic guiding our tests is that a good measure of inter- or intra- cluster connectivity should accurately reflect the simulated graph's structures. We expect measures of intra-cluster connectivity, $\bar{K}_{\mbox{intra}}$ and modularity to increase in step with the simulated graph's intra-cluster connectivity levels, by definition. Meanwhile we expect conductance to display the inverse behavior. We also expect $\bar{K}_{\mbox{inter}}$ to follow the fluctuations of inter-cluster connectivity, by construction.

It should also be mentioned that some authors have used so-called ``ground-truth'' data sets, as benchmarks for clustering algorithm performance (e.g., \cite{bigClam,Moschopoulos2011,YangLesko2012}). These are data sets where the nodes' cluster memberships are known in advance. Our approach is more general, since it is data set independent. Arguably, the fact that an algorithm anecdotally provided accurate clustering on one labeled instance is no guarantee it will perform equally well on another (likely unlabeled) instance. In addition, our experiments provide us with an understanding of each measure's sensitivity and response to graph structure.

\subsection{Experimental Set-up and Results} \label{results}
We experiment with variations in edge probability, both within and between clusters. Here, we slightly modify the procedure to generate inter-cluster edges. In our previous article \cite{PMEtAlWAW18}, we varied the proportion of vertices inside and outside each cluster that shared an edge. Here, we vary edge probabilities. 

Our data generation process consists of modifying intra-cluster and inter-cluster edge probabilities of a planted partition model \cite{FortunatoLong2010} and generating graphs with clusters of varying sizes. We begin with increases in intra-cluster connectivity in steps of 25\%, while maintaining  inter-cluster edge probability at 0\%. For example, in the second column of Table~\ref{deltaIntraInter0}, approximately 25\% of all possible edges within a cluster are added, but nodes only have connections to other nodes within their assigned cluster. Each cluster remains a connected component disconnected from the rest of the graph. 

We conduct these tests with unweighted graphs, but also repeat them with weighted ones. While there are no formal and universally accepted definitions of weighted stochastic block models or planted partition models, we generate data that is consistent with the logic of the planted partition model. In the weighted case, the intra-cluster edge probability also corresponds to edge weight. For example, when edge probability is 25\%, edge  weight is also set to 0.25. All edge weights are between 0 and 1. 

We then examine the effect of inter-cluster connectivity on each measure. We begin with no inter-cluster connectivity and then increase it in steps of 25\%. We increase edge probability between nodes in different clusters in steps of 25\%, while keeping intra-cluster connectivity at 0\%. In other words, clusters are just sets of non-adjacent vertices. In these scenarios, we imagine an algorithm, a very poorly performing one, that groups non-adjacent vertices into clusters with different levels of inter-connection to other clusters but with an intra-cluster connectivity that remains constant at 0\%. Here again, we also repeat our tests on weighted graphs, with edge weights corresponding to the inter-cluster connectivity percentage. Results are shown in Table~\ref{deltaInterIntra0}.

We acknowledge these synthetic networks are unrealistic. Our goal is not to study common network structures. Our goal is to examine the effects of drastic degenerate structures on our quality measures and compare them to the effects on the competing measures. 

Finally, in order to assess our measures' robustness, we repeat all the tests described above, but with the introduction of noise in the connectivity patterns. Noise is introduced in the form of 100\% intra-(inter-) cluster connectivity (edge probability). Results are shown in Table~\ref{deltaIntraInter1} and Table~\ref{deltaInterIntra1}.

\begin{table}[]
\centering
\caption{Varying Intra-Cluster Connectivity, No Noise from Inter-Cluster Connectivity}
\label{deltaIntraInter0}
\begin{tabular}{|r|| l | l| l| l| l |}
\hline
\multicolumn{6}{| l |}{\textbf{Pct Inter = 0, Pct Intra varies}} \\
\hline
{\bf Pct Intra}                      & 0   & 25     & 50     & 75      & 100       \\
\hline
\multicolumn{6}{|c|}{\textbf{UNWEIGHTED}}  \\
\hline                                                          
N                        & 10,048 & 9,725  & 10,374  & 9,490   & 9,700   \\
$\vert C \vert$          & 200    & 200    & 200     & 200     & 200     \\
$\vert E \vert$          & 0      & 77,043 & 173,221 & 224,723 & 313,955 \\
K                        & 0.0000 & 0.0016 & 0.0032  & 0.0050  & 0.0067  \\
$\bar{K}_{\mbox{intra}}$ & 0.0000 & 0.2640 & 0.4995  & 0.7523  & 0.9900  \\
$\bar{K}_{\mbox{inter}}$ & 0.0000 & 0.0000 & 0.0000  & 0.0000  & 0.0000  \\
$\Phi$                   & nan    & 0.0000 & 0.0000  & 0.0000  & 0.0000  \\
Q                        & nan    & 0.9908 & 0.9911  & 0.9906  & 0.9907 \\
\hline
\multicolumn{6}{|c|}{\textbf{WEIGHTED}}                                                               \\
\hline
N                        & 10,048 & 9,725  & 10,374 & 9,490   & 9,700   \\
$\vert C \vert$          & 200    & 200    & 200    & 200     & 200     \\
$\vert E \vert$          & 0      & 19,261 & 86,611 & 168,542 & 313,955 \\
K                        & 0.0000 & 0.0004 & 0.0016 & 0.0037  & 0.0067  \\
$\bar{K}_{\mbox{intra}}$ & 0.0000 & 0.0660 & 0.2497 & 0.5642  & 0.9900  \\
$\bar{K}_{\mbox{inter}}$ & 0.0000 & 0.0000 & 0.0000 & 0.0000  & 0.0000  \\
$\Phi$                   & nan    & 0.0000 & 0.0000 & 0.0000  & 0.0000  \\
Q                        & nan    & 0.9908 & 0.9911 & 0.9906  & 0.9907 \\
\hline
\end{tabular}
\end{table}

\begin{table}[]
\centering
\caption{Varying Inter-Cluster Connectivity, No Noise from Intra-Cluster Connectivity}
\label{deltaInterIntra0}
\begin{tabular}{|r|| l | l| l| l| l |}
\hline
\multicolumn{6}{|l|}{\textbf{Pct Intra = 0, Pct Inter varies}}                                  \\
\hline
\textbf{Pct Inter}       & 0   & 25      & 50        & 75         & 100          \\
\hline
\multicolumn{6}{|c|}{\textbf{UNWEIGHTED}}                                                       \\
\hline
N                        & 10,048 & 10,048     & 10,048     & 10,048     & 10,048     \\
$\vert C \vert$          & 200    & 200        & 200        & 200        & 200        \\
$\vert E \vert$          & 0      & 12,210,800 & 24,864,800 & 37,291,600 & 50,142,500 \\
K                        & 0.0000 & 0.2419     & 0.4926     & 0.7388     & 0.9934     \\
$\bar{K}_{\mbox{intra}}$ & 0.0000 & 0.0000     & 0.0000     & 0.0000     & 0.0000     \\
$\bar{K}_{\mbox{inter}}$ & 0.0000 & 0.2366     & 0.4907     & 0.7336     & 1.0000     \\
$\Phi$                   & nan    & 1.0000     & 1.0000     & 1.0000     & 1.0000     \\
Q                        & nan    & -0.0067    & -0.0067    & -0.0067    & -0.0067  \\
\hline
\multicolumn{6}{|c|}{\textbf{WEIGHTED}}                                                         \\
\hline
N                        & 10,048 & 10,048    & 10,048     & 10,048     & 10,048     \\
$\vert C \vert$          & 200    & 200       & 200        & 200        & 200        \\
$\vert E \vert$          & 0      & 3,052,700 & 12,432,400 & 27,968,700 & 50,142,500 \\
K                        & 0.0000 & 0.0605    & 0.2463     & 0.5541     & 0.9934     \\
$\bar{K}_{\mbox{intra}}$ & 0.0000 & 0.0000    & 0.0000     & 0.0000     & 0.0000     \\
$\bar{K}_{\mbox{inter}}$ & 0.0000 & 0.0592    & 0.2454     & 0.5502     & 1.0000     \\
$\Phi$                   & nan    & 1.0000    & 1.0000     & 1.0000     & 1.0000     \\
Q                        & nan    & -0.0067   & -0.0067    & -0.0067    & -0.0067   \\
\hline
\end{tabular}
\end{table}

\begin{table}[]
\centering
\caption{Varying Intra-Cluster Connectivity, with Noise from Inter-Cluster Connectivity}
\label{deltaIntraInter1}
\begin{tabular}{|r|| l | l| l| l| l |}
\hline
\multicolumn{6}{|l|}{\textbf{Pct Inter = 100, Pct Intra varies}}                                                        \\
\hline
\textbf{Pct Intra}                  & 0          & 25      & 50         & 75         & 100          \\
\hline
\multicolumn{6}{|c|}{\textbf{UNWEIGHTED}}                                                                             \\
\hline
N                        & 10,048     & 9,725      & 10,374     & 9,490      & 9,700      \\
$\vert C \vert$          & 200        & 200        & 200        & 200        & 200        \\
$\vert E \vert$          & 50,142,500 & 47,052,100 & 53,631,600 & 44,950,500 & 47,040,200 \\
K                        & 0.9934     & 0.9951     & 0.9968     & 0.9983     & 1.0000     \\
$\bar{K}_{\mbox{intra}}$ & 0.0000     & 0.2640     & 0.4995     & 0.7523     & 0.9900     \\
$\bar{K}_{\mbox{inter}}$ & 1.0000     & 1.0000     & 1.0000     & 1.0000     & 1.0000     \\
$\Phi$                   & 1.0000     & 0.9974     & 0.9952     & 0.9922     & 0.9899     \\
Q                        & -0.0067    & -0.0050    & -0.0033    & -0.0018    & -0.0001   \\
\hline
\multicolumn{6}{|c|}{\textbf{WEIGHTED}}                                                                               \\
\hline
N                        & 10,048     & 9,725      & 10,374     & 9,490      & 9,700      \\
$\vert C \vert$          & 200        & 200        & 200        & 200        & 200        \\
$\vert E \vert$          & 50,142,500 & 46,994,300 & 53,545,000 & 44,894,300 & 47,040,200 \\
K                        & 0.9934     & 0.9939     & 0.9952     & 0.9971     & 1.0000     \\
$\bar{K}_{\mbox{intra}}$ & 0.0000     & 0.0660     & 0.2497     & 0.5642     & 0.9900     \\
$\bar{K}_{\mbox{inter}}$ & 1.0000     & 1.0000     & 1.0000     & 1.0000     & 1.0000     \\
$\Phi$                   & 1.0000     & 0.9994     & 0.9976     & 0.9941     & 0.9899     \\
Q                        & -0.0067    & -0.0062    & -0.0049    & -0.0030    & -0.0001 \\
\hline
\end{tabular}
\end{table}

\begin{table}[]
\centering
\caption{Varying Inter-Cluster Connectivity, with Noise from Intra-Cluster Connectivity}
\label{deltaInterIntra1}
\begin{tabular}{|r|| l | l| l| l| l |}
\hline
\multicolumn{6}{|l|}{\textbf{Pct Intra = 100, Pct Inter varies}}                                      \\
\hline
\textbf{Pct Inter}       & 0       & 25      & 50        & 75         & 100          \\
\hline
\multicolumn{6}{|c|}{\textbf{UNWEIGHTED}}                                                           \\
\hline
N                        & 9,700   & 9,700      & 9,700      & 9,700      & 9,700      \\
$\vert C \vert$          & 200     & 200        & 200        & 200        & 200        \\
$\vert E \vert$          & 313,955 & 11,612,400 & 23,403,200 & 34,964,100 & 47,040,200 \\
K                        & 0.0067  & 0.2469     & 0.4975     & 0.7433     & 1.0000     \\
$\bar{K}_{\mbox{intra}}$ & 0.9900  & 0.9900     & 0.9900     & 0.9900     & 0.9900     \\
$\bar{K}_{\mbox{inter}}$ & 0.0000  & 0.2294     & 0.4842     & 0.7271     & 1.0000     \\
$\Phi$                   & 0.0000  & 0.9595     & 0.9798     & 0.9864     & 0.9899     \\
Q                        & 0.9907  & 0.0202     & 0.0066     & 0.0022     & -0.0001   \\
\hline
\multicolumn{6}{|c|}{\textbf{WEIGHTED}}                                                             \\
\hline
N                        & 9,700   & 9,700     & 9,700      & 9,700      & 9,700      \\
$\vert C \vert$          & 200     & 200       & 200        & 200        & 200        \\
$\vert E \vert$          & 313,955 & 3,138,570 & 11,858,600 & 26,301,500 & 47,040,200 \\
K                        & 0.0067  & 0.0667    & 0.2521     & 0.5591     & 1.0000     \\
$\bar{K}_{\mbox{intra}}$ & 0.9900  & 0.9900    & 0.9900     & 0.9900     & 0.9900     \\
$\bar{K}_{\mbox{inter}}$ & 0.0000  & 0.0574    & 0.2421     & 0.5453     & 1.0000     \\
$\Phi$                   & 0.0000  & 0.8556    & 0.9603     & 0.9820     & 0.9899     \\
Q                        & 0.9907  & 0.0930    & 0.0196     & 0.0051     & -0.0001 \\
\hline
\end{tabular}
\end{table}

\subsection{Interpretation of Empirical Comparisons}
As shown in Section~\ref{results}, our Kappas behave exactly as expected, even when subjected to noise. In all instances where the labeling of clusters reflects a good partition, the inequalities $\bar{K}_{\mbox{intra}} > K > \bar{K}_{\mbox{inter}}$ hold and they do not not hold in instances where the partition reflects poor clustering. For example, in Table~\ref{deltaIntraInter1}, all instances are cases of poor clustering. Similarly, in Table~\ref{deltaInterIntra1}, instances where the percentage of inter-cluster connectivity is below 75\% are examples of good clustering. Decimal mismatches with the expected edge probabilities are due to rounding in the sampling procedure. In contrast, we note modularity and conductance display very counterintuitive behaviors. 

The case of modularity is of particular interest, given it is the most widely used clustering quality measure. To illustrate the breakdown of modularity observed in Tables~\ref{deltaIntraInter0}-\ref{deltaInterIntra1}, we use a hypothetical example of an unweighted graph with $k$ clusters all containing the same number of vertices $(\vert v_i \vert = v, \, \forall i)$ and with equal inter- and intra-cluster edge probability for all individual clusters and cluster pairs. 

\subsubsection{Case 1 (Table~\ref{deltaIntraInter0}), varying intra-cluster edge probability, with inter-cluster edge probability of zero:} 
In this case, inter-cluster edge probability is zero and intra-cluster edge probability is denoted by $p \; (\ge 0)$. In this case, modularity is approximated as 
\begin{eqnarray*}
Q &=& \sum_{i=1}^{k} \left( \underbrace{ e_{ii} - a_i^2 }_{ q_i } \right)  \\
&\approx& k \times \left[ e_{11} - a_1^2 \right]  \, .
\end{eqnarray*}
If $p=0$, modularity is undefined, because $m=0$. If we let $p >0$, since we have only intra-cluster edges, a constant edge probability and number of vertices for all clusters, all $e_{ii}$ and all $a_i$ are not only equal within a given cluster. They also take on the same approximate value, across all clusters:
\begin{eqnarray*}
e_{ii} = a_i &=& \frac{1}{2m} \times 2 \times \vert E_{ii} \vert \quad \forall i \in C \\
\vert E_{ii} \vert &\approx& p \times 0.5 \times v \times \left( v - 1 \right) \approx \frac{m}{k} \quad \forall i \in C \\
\sum_{i=1}^k \vert E_{ii} \vert &=& m\, .
\end{eqnarray*}
As soon as $p$ increases to any non-zero value, modularity reaches its maximum and remains unaffected by variations in connectivity, as shown here.
\begin{eqnarray*} 
Q &\approx& k \times  \left[ \frac{1}{2m} \frac{2m}{k} - \frac{1}{4m^2} \frac{4m^2}{k^2} \right] \\
&\approx& 1 - \frac{1}{k} \approx 1 \quad (\text{if $k$ is large})
\end{eqnarray*}

\subsubsection{Case 2 (Table~\ref{deltaInterIntra0}), varying inter-cluster edge probability, intra-cluster edge probability of zero:} 
Here, since there are no intra-cluster edges, we have $e_{ii} = 0$ for each cluster $i$. Since the inter-cluster probability is uniform across all cluster pairs ($p$) and intra-cluster edge probability is zero, the approximate number of connections (edge stubs) originating from each cluster is $ s_i \approx \left( p \times v \times (k-1) \times v \right) \approx \frac{2m}{k}$. The approximate number of stubs is equal to the probability of inter-cluster edge multiplied by the total number of possible inter-cluster connections originating in cluster $i$ and connecting to vertices in all other remaining $(k-1)$ clusters. Here, we note there are $v$ vertices in each cluster $i$ and $(k-1)v $ vertices in the remaining clusters. Summed over the graph, it corresponds to two times the number of inter-cluster edges originating in each cluster:
\[
\sum_{i=1}^k s_i = 2m, \;
s_i \approx \frac{2m}{k} \; .
\]
Modularity with no intra-cluster edges but a probability of inter-cluster edge $p$ then becomes approximately:
\begin{eqnarray*}
Q &\approx& k \times \left[ 0 -  \frac{1}{4m^2}\left( p \times v \times (k-1) \times v  \right)^2 \right] \\
&\approx& k \times \left[ - \frac{1}{4m^2} \left( \underbrace{ p \times v \times (k-1) \times v }_{s_i}\right)^2 \right]  \\
&\approx& k \times \left[ - \frac{1}{4m^2} \left( \frac{2m}{k} \right)^2 \right] = -\frac{1}{k} \\
&\approx& 0 \quad (\text{if $k$ is large}) \, .
\end{eqnarray*}
Here again, modularity remains fixed, this time around zero, and unaffected by connectivity patterns.

\subsubsection{Case 3 (Table~\ref{deltaIntraInter1}), varying intra-cluster edge probability, inter-cluster edge probability of 1:} 
In Table~\ref{deltaIntraInter1}, we begin our experiment with an intra-cluster edge probability of zero. This situation takes us back to the hypothetical example described in Case 2. In that example, modularity was approximately equal to zero. 

Increasing intra-cluster edge probability has only a very minor effect on modularity because there are far more inter-cluster edges than intra-cluster ones. At each step, the denominator, number of edges, goes up minimally as the number of intra-cluster edges increases. The numerator also increases, but it remains smaller than the denominator. For example, increasing intra-cluster edge probability from zero to some $p$ ($>0$) results in the changes below. In this example, $m$ is the number of edges when intra-cluster edge probability is zero.
\begin{align*}
& e_{ii}  = \frac{p \times v \times \left(v - 1 \right)}{2m + p \times v \times \left(v - 1 \right)} \approx \epsilon_e \\
& a_i^2 = \frac{\left( \frac{2m}{k} + p \times v \times \left( v - 1 \right) \right)^2}{\left( 2m + p \times v \times \left( v - 1 \right) \right)^2} \approx \epsilon_a \\
& p \times v \times \left( v - 1 \right)^2 \ll 2m \\
& \epsilon_e - \epsilon_a = \epsilon  \:  \sim 0 \\
& k \times \epsilon \sim 0
\end{align*}
As a result, modularity only increases in a trivial manner and remains at a value of approximately zero.

\subsubsection{Case 4 (Table~\ref{deltaInterIntra1}), varying inter-cluster edge probability, intra-cluster edge probability of 1:}
In Table~\ref{deltaInterIntra1}, modularity begins at a value approximately equal to one. This graph structure is the same as the last example in Table~\ref{deltaIntraInter0}, where intra-cluster edge probability is equal to one and inter-cluster edge probability is equal to zero. 

When inter-cluster edge probability increases to $p > 0$, the numerator in the $e_{ii}$ remain constant, since intra-cluster edge probability is one. However the denominator increases dramatically and decreases the $e_{ii}$ very drastically. Let $m$ denote the number of edges with inter-cluster edge probability of zero. When inter-cluster edge probability increases to $p > 0$, the number of edges becomes
\[
\vert E \vert = m + k \times \left( \frac{1}{2} \times p  \times v \times \left[ (k-1) \times v \right] \right) \, . 
\]
In this case, each cluster adds $\frac{1}{2} \times p \times v \times (k-1) \times v$ edges. This increase in the number of edges causes $e_{ii}$ to collapse, as soon as $p > 0$.
\begin{align*}
& e_{ii}  = \frac{v \times \left( v - 1 \right)}{2m + \left( k \times p \times v \times \left[ \left( k -1 \right) \times v  \right] \right)} \approx 0 \\
\end{align*}
Because,
\begin{align*}
& v \times \left( v - 1 \right) \ll 2m + \left(k \times p \times v \times \left[ \left( k -1 \right) \times v  \right] \right) \\
\end{align*}
Meanwhile, when inter-cluster edge probability increases to $p > 0$, $a_i^2$ increases slightly. 
\[
a_i^2 = \frac{ \left[ v \times \left( v - 1 \right) + p \times v \times \left( k - 1 \right) \times v \right]^2 }{ \left[ 2m + k \times p \times v \times \left( k - 1 \right) \times v \right]^2} = \epsilon ( \gtrsim e_{ii} )
\]

\subsubsection{Link to Axioms}
The axioms listed in Section~\ref{axioms}, which define the properties of good clustering quality functions, may offer some clues to interpret modularity's counter-intuitive results (Tables~\ref{deltaIntraInter0}-\ref{deltaInterIntra1}). While modularity does not violate the axioms of richness, monotonicity and consistent improvement and perfectness, it only barely meets them. It also appears to violate continuity, in our experiments.

Richness essentially means the optimum must be achievable through a re-arrangement of edges. Although not defined as such, we would also expect that a ``rich'' function's minimum also be achievable, in routine cases, not just extremely rare degenerate trivial cases. However, the minimum for modularity, $Q = -\frac{1}{2}$, is only achievable in very rare degenerate and trivially identifiable cases. This fact is documented by Brandes et al. \cite{modBrandes2007}. Van Mieghen et al. also corroborate this claim about the lower bound, when they state { \it ``In conclusion, the modularity of any graph is never smaller than $-\frac{1}{2}$ , and this minimum is obtained for the complete bipartite graph''} \cite{modAssort}. Empirically, this weak realization of richness can be seen in all cases where intra-cluster edge probability is zero. Arguably, a good quality function should be at its minimum, when intra-cluster edge probability is zero. Unfortunately, in all of our tests, modularity never reaches its minimum, regardless of how poor the clustering. These cases can be seen in the first column of Table~\ref{deltaIntraInter0} where modularity (Q) is undefined, in Table~\ref{deltaInterIntra0} where Q is either undefined or roughly equal to zero and in the first column of Table~\ref{deltaIntraInter1}.

Once again, while it does not violate the axiom of monotonicity and consistent improvement modularity only meets it weakly. Monotonicity and consistent improvement mean that an increase in edge density within clusters or decrease in inter-cluster edge density should not decrease the quality function. While modularity does not decrease with improvements in clustering, it rapidly meets its maximum after trivial improvements and then fails to accurately reflect any further improvements. This inconsistent behavior can be seen in Table~\ref{deltaIntraInter0} and Table~\ref{deltaIntraInter1} where modularity meets its maximum of one, with an intra-cluster edge probability of $25\%$ and then fails to reflect any additional increase in intra-cluster edge probability.

Continuity is an axiom that is clearly violated by modularity. According to Van Laarhoven and Marchiori, ``A quality function Q is continuous if a small change in the graph leads to a small change in the quality''. However, as mentioned previously, modularity spikes from a value of zero to its maximum of one, after only a modest improvement in clustering, an increase of $25\%$ intra-cluster edge probability. 

Perfectness is another axiom that is only barely met by modularity. While a set of disjoint complete graphs does indeed get a ``perfect'' modularity score, the reverse is not true. We would expect a ``perfect'' clustering quality function to reach its minimum when clustered vertices form a complete k-partite graph (where $k>2$). Unfortunately, modularity does not reach its minimum in such cases. In our experiments, modularity of such poorly clustered graphs is shown in the fifth column of Table~\ref{deltaInterIntra0} and the first column of Table~\ref{deltaIntraInter1}.

\section{Illustrative Example: Comparing Clustering Quality of the Louvain and Asynchronous Label Propagation Algorithms}
\begin{figure}[H] 
\centering
\subfloat[EUC, 35 runs, 20 clust]{ \includegraphics[width = 0.3\textwidth]{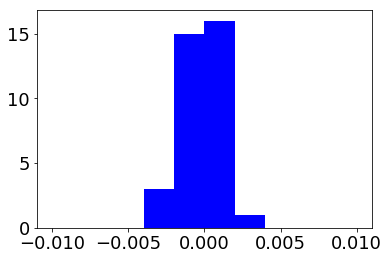} }
\subfloat[EUC, 35 runs, 27 clust]{ \includegraphics[width = 0.3\textwidth]{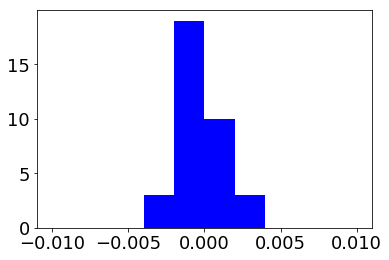} } 
\caption{Null Distributions of $\gamma$, EUC Graph} \label{null2}
\end{figure}

To illustrate the application of each step of our work, we test the Louvain \cite{Louvain2008} and the Asynchronous Label Propagation (ALP) \cite{asynLabProp2007} algorithms, as implemented in the Networkx library's \cite{Networkx} ``Communities'' module. We use these algorithms to cluster the SNAP ``email-Eu-core network'' graph \cite{SNAP2007,SNAP2017}, which we converted into an undirected graph with no self-loops (EUC).
\begin{table}[H] 
\begin{center}
\caption {Key Characteristics of the Graph and Simulations of the Null Hypotheses } \label{sims2}
\begin{tabular}{| c| c| c| c| c| c| c|}
\hline
{\bf Graph} & {\bf Num Vertices} & {\bf Num Edges} & {\bf Num Clusters} &{\bf Num Runs} & {\bf Mean $\gamma$} & {\bf Std $\gamma$}\\
\hline
EUC    & 1,005 & 16,064 & 20 & 35  & -0.0001	 & 0.0013   \\
EUC    & 1,005 & 16,064 & 27 & 35   & -0.0002  & 0.0014   \\
\hline
\end{tabular}
\end{center}
\end{table}
To obtain standard errors and to verify our claims about the null distribution in small sample (number of clusters) cases, once more, we simulate 35 instances of random cluster assignments. The number of clusters used to simulate the null distribution is the same as the number of clusters identified by each of the two clustering algorithms. Sample statistics of the null and graph characteristics are shown in Table~\ref{sims2}. Corresponding histograms are shown in Figure~\ref{null2}. Full statistical test results are shown in Table~\ref{empres}.

\begin{table}[H]
\begin{center}
\caption {Clustering Results: Test Statistics} \label{empres}
\begin{tabular}{| c| c| c| c| c| c| c| c| c|}
\hline
{ \bf Algorithm} & {\bf Num Clusters} & {\bf $\bar{K}_{\mbox{intra}}$} & {\bf $\bar{K}_{\mbox{inter}}$} & {\bf K} & {\bf $\gamma$} & {\bf t-stat} & {\bf df} & {\bf p-value} \\
\hline
ALP   & 20  &   0.0017 & 0.0000 &  0.0318 & 0.0017 & 1.3077 & 34 & $\approx$ 0.1\\
Louvain & 27 & 0.0513 & 0.0011 & 0.0318  & 0.0502 & 35.8571 & 34 & $\approx$ 0.0\\
\hline
\end{tabular}
\end{center}
\end{table}

\subsection{Analysis}
In the case of the Louvain algorithm, we begin by observing that the inequalities $\bar{K}_{\mbox{intra}} > K > \bar{K}_{\mbox{inter}}$ hold numerically. To pursue our analysis with the second necessary condition, we first verify that the null distribution is centered at zero and roughly symmetric. We then compute our statistical test and observe the hypothesis that $\bar{K}_{\mbox{intra}} = \bar{K}_{\mbox{inter}}$ is rejected. We conclude the clustering returned by the  Louvain algorithm is of good quality and statistically significant. 

Meanwhile, for the ALP algorithm, we see that the inequalities $\bar{K}_{\mbox{intra}} > K > \bar{K}_{\mbox{inter}}$ do not hold numerically. In fact, mean intra-cluster density is lower than global density, $\bar{K}_{\mbox{intra}} < K$. Our analysis could end here, with a conclusion that the ALP algorithm offers a very poor clustering. 

Notwithstanding this violation of our first necessary condition, we pursue our analysis for illustration purpose. Our significance test reveals the null hypothesis that $\bar{K}_{\mbox{intra}} = \bar{K}_{\mbox{inter}}$ is not rejected, at a confidence level of approximately $0.1$. For these reasons, we conclude the clustering returned by the ALP algorithm is of poor quality.

Finally, although it is unnecessary in this specific case, we apply our two-algorithm heuristic test. We examine the difference \[ p_{\text{\tiny ALP}} - p_{\text{\tiny Louvain}} \approx 0.10 \; (\gg 0) \] and obtain further evidence the Louvain algorithm identifies more meaningful clusters than the ALP algorithm.

The case of the ALP algorithm also illustrates the value of the standard error estimation technique in our modified t-test. The ALP algorithm lumps $986$ of the 1,005 vertices into one single cluster. All remaining clusters, $19$ of them are isolated vertices which have no edges connecting them to the rest of the graph. In this specific case, the standard error for the mean inter-cluster density, $\bar{K}_{\mbox{inter}}$, can only be computed through our Monte-Carlo technique. Because there are no inter-cluster edges in this clustering, the standard deviation for inter-cluster density cannot be estimated. Consequently it would be impossible to compute the standard error for $\bar{K}_{\mbox{inter}}$ in the usual way. Fortunately, our modified t-test allows us to overcome this obstacle and compute a t-statistic and p-value.

\section{Conclusion}
We described a new set of statistically-rooted clustering quality measures that allow formal clustering quality assessments and comparisons of clustering algorithm performances. Our measures are shown to be more robust than the commonly used modularity and conductance. In particular, our measures appear to be more responsive to cluster labeling and less sensitive to sample size and breakdowns during numerical stress testing. We also adapted Student's two-sample t-test to circumvent any possible correlation and degeneracies.

Future work will be focused on developing quality measures for clusterings with overlapping clusters. Overlapping clusters offer a more realistic model of most real-world networks. Unfortunately, quality measures to evaluate algorithms that identify overlapping clusters are sparse and a thorough examination is still required. Some authors have extended modularity to overlapping cluster quality measurements (e.g., \cite{modOverlap2010,modOverlap2015}). However, a thorough evaluation of it is still required. It is highly likely that the overlapping cluster extension suffers from many of the same shortcomings as its non-overlapping parent measure, including not meeting the axioms defining good clustering functions.

\section*{Funding}
PM was supported by Mitacs-Accelerate PhD award IT05806.

\section*{Acknowledgements}
PM thanks Leonidas Pitsoulis, Liudmila Prokhorenkova, Andrei M. Raigorodskii, Cris Moore, Aaron Clauset and Mark Newman for their helpful comments. 

\bibliography{consolidatedBib}

\end{document}